\documentclass[twocolumn]{aastex631}

\usepackage{amssymb,amsmath,multirow,bm}

\usepackage{xcolor}
\definecolor{grn}{RGB}{25,25,120}
\definecolor{red}{RGB}{168,0,0}

\usepackage{booktabs}
\usepackage{CJK}
\newcommand{\full}{${\tt full\_1}$}
\newcommand{\vmax}{${V_{\rm max}}$}
\newcommand{\dcvmax}{${V^{\rm DC}_{\rm max}}$}
\newcommand{\sdcvmax}{${V^{\rm SDC}_{\rm max}}$}

\newcommand{\mpch}{{h^{-1}\rm Mpc}}

\newcommand{\kms}{{\rm km}\,\mathrm{s}^{-1}}

\defcitealias{2006MNRAS.368...21L}{Li06}
\defcitealias{2020RAA....20...54Y}{Paper I}

\shorttitle{The smoothed \vmax\ method}
\shortauthors{Yang et al.}
\graphicspath{{./}{figures/}}

\begin{document}
\begin{CJK}{UTF8}{gkai}

\title{Toward Accurate Measurement of Property-Dependent Galaxy Clustering:\\
II. Tests of the Smoothed Density-corrected \vmax\ Method}

\correspondingauthor{Lei Yang}
\email{leiyangastro@ynu.edu.cn }

\author[0000-0001-8297-0868]{Lei Yang (杨蕾)}
\affiliation{South-Western Institute for Astronomy Research, Yunnan University \\ 
Kunming, Yunnan 650500, China}

\author{Zhigang Li (李志刚)}
\affil{College of Physics and Electronic Engineering, Nanyang Normal University\\ Nanyang, Henan, 473061, China}

\begin{abstract}

We present a smoothed density-corrected \vmax\ technique for building a random catalog 
for property-dependent galaxy clustering estimation. This approach is essentially based on 
the density-corrected \vmax\ method of \cite{2011MNRAS.416..739C}, 
with three improvements to the original method. To validate the improved method, 
we generate two sets of flux-limited samples from two independent mock catalogs 
with different $k+e$ corrections. By comparing the two-point correlation functions, 
our results demonstrate that the random catalog created by 
the smoothed density-corrected \vmax\ approach provides a more accurate and precise measurement 
for both sets of mock samples than the commonly used \vmax\ method and redshift shuffled method. 
For flux-limited samples and color-dependent subsamples, the accuracy for the projected correlation function 
is well constrained within $1\%$ on the scale of $0.07\mpch - 30\mpch$. 
The accuracy of the redshift-space correlation function is less than $2\%$ as well. 
Currently, it is the only approach that holds promise for achieving the goal of high-accuracy 
clustering measures for next-generation surveys.

\end{abstract}


\section{Introduction} \label{sec:intro}

Over the last couple of decades, the successful observation of galaxy redshift surveys 
(e.g., Two Degree Field Galaxy Redshift Survey (2dFGRS) \citealt{2003astro.ph..6581C};
the Sloan Digital Sky Survey (SDSS) \citealt{2000AJ....120.1579Y}; the Baryon Oscillation 
SpectroscopicSurvey (BOSS) \citealt{2011AJ....142...72E}; the VIMOS Public Extragalactic 
Redshift Survey (VIPERS) \citealt{2012PASP..124.1232G}) have enabled significant progress toward 
our understanding of galaxy formation and evolution 
\citep{2003MNRAS.344..847M,2006ApJS..167....1B,2011MNRAS.413..101G,2018ApJ...858...30G,
2021MNRAS.505.5117Z}, the galaxy-halo connection \citep{1998ApJ...494....1J,
2003MNRAS.339.1057Y,
2008ApJ...676..248Y,2012ApJ...752...41Y,2005ApJ...633..791Z,2009ApJ...707..554Z,
2004MNRAS.353..189V,
2021MNRAS.504.4667A,2018ARA&A..56..435W,2019MNRAS.488.3143B}, 
and the nature of gravity and dark energy 
( \citealt{,2001Natur.410..169P,2013PhR...530...87W,2013MNRAS.429.1514S,
2021PhRvD.103h3533A} and reference therein). 
In the upcoming years, next-generation surveys, 
such as the Dark Energy Spectroscopic Instrument (DESI; \citealt{2013arXiv1308.0847L,
2016arXiv161100036D,2016arXiv161100037D}), the Legacy Survey of Space and Time 
(LSST; \citealt{2012arXiv1211.0310L}), the space mission Euclid 
\citep{2013LRR....16....6A} and CSST \citep{2018MNRAS.480.2178C,2019ApJ...883..203G},
will map the 3D galaxy distribution in an unprecedentedly volume, leading to about an order of 
magnitude more extragalactic spectroscopic redshifts than those that SDSS, BOSS and eBOSS have 
achieved \citep{2021arXiv210613120Z,2022arXiv220212911Y,2022arXiv220808518M,2022arXiv220903585S}. 
Massive amounts of data from deeper in the sky will provide new insights into the physics of 
galaxy formation, as well as the nature of dark matter and dark energy \citep{2022arXiv220808512H}. 
Galaxy two-point statistics, being one of the most fundamental tools, 
will continue to play a crucial role in future data analysis \citep{2022arXiv220307491V,2022arXiv220307946A}, 
as they have in the past \citep{2011ApJ...736...59Z,2013MNRAS.432..743N,2014ApJ...784..128S,
2014MNRAS.439.3504S,2014MNRAS.441.2398G,2016A&A...594A..13P,2018ApJ...861..137S}. 
Due to different systematics, it is still difficult to reliably measure small-scale 
property-dependent galaxy clustering at the present time. 
These systematics include redshift-dependent completeness, the missing galaxies 
in observations \citep{2016MNRAS.455.1553R,2017MNRAS.472.1106B,
2020MNRAS.495.1511B}, the incorrect estimation of the radial selection model 
\citep{2012MNRAS.424..564R,2020RAA....20...54Y}, among others 
\citep{2021A&A...646A..40B,2021MNRAS.507.3187F,2021MNRAS.506.2503M}.
Fortunately, the coming big data will considerably reduce random errors 
in clustering determination, but to reach the high accuracy of clustering analysis 
required by the next generation surveys, we must eliminate systematic errors 
in measurement \citep{2014MNRAS.443.1065B,2016MNRAS.455.1553R,2021MNRAS.503.3510G,
2021MNRAS.506.4667D}. In this study, the systematic bias produced by the radial selection 
model is investigated in greater detail.

To measure the galaxy two-point correlation function (hereafter 2PCF), 
we must build a random catalog with the same angular and radial selection 
functions as the observed sample, but with a random distribution in the observed space 
\citep{1983ApJ...267..465D,1993ApJ...417...19H}. The angular selection function is easy to obtain from 
observation, but the radial selection function is difficult to estimate accurately. 
As the sample has a fixed number density and the redshift distribution of 
a random catalog is straightforward to construct \citep{2004PhRvD..69j3501T}, 
previous works often use a volume-limited sample for clustering analysis 
\citep{2002MNRAS.332..827N,2002ApJ...571..172Z,
2005ApJ...630....1Z,2011ApJ...736...59Z,2011ApJ...726...13M,2016ApJ...833..241S,
2018A&A...610A..59M}. However,
due to the need of excluding a substantial number of galaxies, 
the statistical precision of the clustering measurement is reduced 
\citep{2005ApJ...630....1Z,2016MNRAS.460.3647X}. 
Alternatively, a flux-limited sample may optimize the utilization of 
observed galaxies, but since its radial selection function $\phi(z)$ 
changes with redshift, it is not easy to build the redshifts of random galaxies for 
a flux-limited sample unless we know the galaxy luminosity function 
(LF) $\Phi(M_{\rm r})$ \citep{2015MNRAS.451.1540L,2021arXiv210906136K}.

The radial selection function for the flux-limited sample has been recovered using 
a number of ways. For instance, the smooth spline fit approach utilizes 
a {\it spline} model to fit the redshift distribution of 
a galaxy sample \citep{2010MNRAS.404...60R,2017MNRAS.472.2869W}. 
The \vmax\ method populates random galaxies within the maximum viewable volume 
of a real galaxy, which is dependent on the galaxy's observational limitations.
The redshift {\it shuffled} technique is a commonly employed alternative 
\citep{2013ApJ...767..122G,2015MNRAS.454.1161Z,2021SCPMA..6489811W}. 
This approach chooses redshifts at random from the real galaxy sample and assigns 
them to the random galaxy catalog. 
Through clustering analysis of the VIPERS data, \cite{2013A&A...557A..54D} showed 
that the spline fit approach underestimates the predicted 2PCF in comparison to 
the \vmax\ method, particularly on scales larger than $3~\mpch$.
In the BOSS systematics investigation, \cite{2012MNRAS.424..564R} revealed 
that the shuffled technique had a minor bias in BAO measurement compared to 
the spline fit method \citep{2015MNRAS.449..835R}. 
However, \cite{2019JCAP...08..036D} demonstrated that 
the shuffled approach suffers from the {\it integral constraint} effect when measuring 
the power spectrum. Using mocks from a high-resolution simulation, 
\citet{2020RAA....20...54Y} (hereafter \citetalias{2020RAA....20...54Y}) 
found that both the redshift shuffled technique and \vmax\ method 
underestimate galaxy clustering by 30\% and 20\%, respectively, 
on scales $\gtrsim ~ 10\mpch$ for flux-limited samples.
Consequently, as long as we continue to use the aforementioned 
radial selection methods to construct the redshifts for random catalogs for 
a flux-limited sample, our clustering measurement will contain 
an unavoidable systematic deviation from the true galaxy clustering.

\cite{2011MNRAS.416..739C} proposes a density-corrected \vmax\ technique 
for concurrently estimating LF and generating a random catalog for a flux-limited sample. 
Unlike the conventional \vmax\ method, this technique can successfully eliminate 
density fluctuations. In \cite{2011MNRAS.416..739C}, they examine the radial distribution 
of random galaxies, which is in excellent agreement with the input galaxy sample. 
This method has been employed to determine property-dependent galaxy clustering 
\citep{2015MNRAS.454.2120F} and clustering analysis 
\citep{2017A&A...608A..44D,2017A&A...604A..33P,
2018MNRAS.474.3435L,2021A&A...646A.147J}. 
However, its clustering measurement performance has not been assessed. 
The purpose of this study is to test the \cite{2011MNRAS.416..739C} technique 
for clustering measurements using mock data.
In addition, some modifications are made to the original approach 
in order to improve its measurement accuracy.

This paper is structured as follows. In Section~\ref{sec:method}, 
we review the \cite{2011MNRAS.416..739C} method and introduce 
the smoothed density-corrected \vmax\ method. 
The construction of mock galaxy catalogs is detailed in 
Section~\ref{sec:mocks}. We present the testing results of the correlation functions 
in Section~\ref{sec:tests}. In Section~\ref{sec:disc}, we assess the smoothed 
density-corrected \vmax\ method and discuss the potential sources of uncertainty 
in estimates. We conclude the paper in Section~\ref{sec:concls}.


\section{Smoothed Density-corrected \vmax\ Method} \label{sec:method}

To address the difficulty of recovering the radial selection function of 
a property-dependent galaxy sample, \cite{2011MNRAS.416..739C} 
 developed a density-corrected \vmax\ approach for galaxy clustering estimates.
This section starts with a briefly overview of the \cite{2011MNRAS.416..739C} technique. 
Following that, we detail the improvements to the original \cite{2011MNRAS.416..739C} 
methodology, which we call the smoothed density-corrected \vmax\ method.

\subsection{\cite{2011MNRAS.416..739C} Method} \label{sec:colemethod}

On the basis of the standard \vmax\ approach, \cite{2011MNRAS.416..739C} 
presented a weighted \vmax\ method based on a joint stepwise maximum 
likelihood method, which effectively eliminates the influence of density fluctuation.
In this method, a density-weighted maximum volume 
\dcvmax\ \footnote{See Equations (11) and (16) 
in \cite{2011MNRAS.416..739C}.} is defined, which is the normal \vmax\ weighted 
by the estimated galaxy overdensities $\Delta(z)$ and 
the LF density evolution $P(z)$.
They further define a weight as 
\begin{equation}
w_{\alpha}\equiv  \frac{V_{\alpha, \rm max}}{V^{\rm DC}_{\alpha,\rm max} + \mu V_{\alpha,\rm max}},
\end{equation} 
where $V_{\alpha, \rm max}$ and $V^{\rm DC}_{\alpha, \rm max}$ are the normal \vmax\ and 
density-corrected \vmax\ for the $\alpha$th galaxy in the observed sample.
$\mu$ is a Lagrange multiplier providing constraints with 
$\langle \frac{V_{\alpha, \rm max}}{V^{\rm DC}_{\alpha,\rm max}+\mu V_{\alpha,\rm max}}\rangle =1$ 
when estimating LF for the galaxy sample.
Lastly, a random catalog can be created by replicating individual galaxies 
$n_{\alpha}=nw_{\alpha}$ times and distributing them at random across 
the $V_{\alpha, \rm max}$ volume.
Note that, unlike the standard \vmax\ approach, $n_{\alpha}$ is no longer 
the same for all galaxies and the selection rate of random galaxies is adjusted by 
weight $w_{\alpha}$. 
The brightness of the galaxy may be over- or underrepresented in the 
observed sample as a result of the density variation in the \vmax\ volume 
being appropriately compensated by the weight $w_{\alpha}$. 
By comparing the output redshift distribution to that of the input galaxy sample, 
\cite{2011MNRAS.416..739C} proved that the random catalog created by 
this density-weighted \vmax\ technique could recover the genuine galaxy selection function. 
While this approach has not yet been tested on galaxy clustering using mock galaxy catalog, 
it remains to be validated using mocks.

\subsection{Smoothed Density-corrected \vmax\ Method} \label{sec:sdcVmethod}

Before testing the \cite{2011MNRAS.416..739C} method, we perform three 
modifications to the original public code~\footnote{\url{http://astro.dur.ac.uk/~cole/random_cats/}}.
The original algorithm is only applicable to galaxy sample with a single faint flux cut, 
but by adding a $z_{\rm min}$ estimate, our first update makes the code applicable 
to a generic double flux-cut sample \footnote{This modification primarily changes 
the step-function $S$ from $S(L^{\rm min} |L)$ to $S(L^{\rm min}, L^{\rm max} | L)$ in 
equation(5) and the lower limit of \vmax\ integration in equation (11) and (39) 
in \cite{2011MNRAS.416..739C}.}.
The maximum(minimum) redshifts $z_{\rm max(min)}$ in our updated code is same as 
 \citetalias{2020RAA....20...54Y} and are determined as follows: 
 \begin{eqnarray}\label{eq:zz1}
   z_{\rm max}&=&\mathtt{min}[z_{\rm mag, max},~z_{\rm sample, max}],\\
   z_{\rm min}&=&\mathtt{max}[z_{\rm mag, min},~z_{\rm sample, min}],
   \label{eq:zz2}
  \end{eqnarray}
where $z_{\rm sample, max(min)}$ is the redshift limits of 
the galaxy sample, and $z_{\rm mag, max(min)}$ is derived by 
\begin{eqnarray}\label{eq:mm1}
m_{\rm faint}&=&M+{\rm DM}(z_{\rm mag, max})+k(z)-E(z),\\ 
m_{\rm bright}&=&M+{\rm DM}(z_{\rm mag, min})+k(z)-E(z),
\label{eq:mm2}
  \end{eqnarray}
where the flux limits are set by apparent magnitude 
$m_{\rm bright(faint)}$,  $M$ is the absolute 
magnitude, the distance modulus is 
${\rm DM}=5{\rm log}_{10}(d_{\rm L})+25-5{\rm log}_{10}h$, 
$k(z)$ is the $k$ correction, and $E(z)$ is the luminosity evolution correction 
($e$ correction). 
Our second code improvement is the $k$ correction.
In the original code, the $k$ correction is performed for all galaxies 
depending on the input function $k(z)$, which hinders the method's ability 
to apply to a real galaxy sample whose $k$ correction is dependent not just 
on redshift but also on galaxy properties (e.g., color). We modify the code to take a 
$k$(z,color) model as input, allowing $k$ correction to be conducted 
on individual galaxies based on their redshifts and colors. 
This makes the technique more applicable to observable galaxies. 
Following the aforementioned modifications, the output cloned random catalog 
from the updated algorithm is basically consistent with the genuine radial distribution of 
the galaxy number density $n_{\rm true}(z)$. 
However, there are small fluctuations in the output radial distribution that have 
a considerable influence on the final clustering estimate. 
Our final modification to the algorithm is to smooth the radial distribution 
of the output cloned random galaxies. In the smooth procedure, we begin by 
generating a histogram of comoving distance $d$ for the random galaxies. 
We set a bin size of $\Delta d=5 \mpch$, and $N(d)_{\rm hist}$ represents 
the number of random galaxies in each bin. 
Second, we employ a convolution operator to smooth the histogram as 
$N^{\rm s}_{\rm hist}=[N_{\rm hist}\ast \Delta_{\rm smooth}]$, where 
$\Delta_{\rm smooth}=5$ is the smoothed box size in 1D and $N^{\rm s}_{\rm hist}$ 
is the smoothed radial distribution of random galaxies. 
Final redshifts for random galaxies are generated based on the profile of 
$N^{\rm s}_{\rm hist}$ that has been smoothed. In Section~\ref{sec:comparison}, 
we will observe that our modifications enhance the clustering measurement 
accuracy significantly.

\cite{2015MNRAS.454.2120F} recently developed the \cite{2011MNRAS.416..739C} 
technique to quantify the property-dependent galaxy clustering of GAMA II data 
\citep{2011MNRAS.413..971D,2015MNRAS.452.2087L}.
They found that the \cite{2011MNRAS.416..739C} technique yields 
a redshift distribution that is too broad for cloned random galaxies, 
which may be the result of luminosity evolution.
To mitigate this unanticipated impact, \cite{2015MNRAS.454.2120F} 
developed a Gaussian window function to restrict the redshift distribution 
of the cloned galaxies. In the first place, the mock galaxy catalogs that we construct 
in this study resemble the low redshift SDSS data, as opposed to the GAMA data, 
which encompass a relatively broad redshift range of 0$\sim$0.5. In our mock galaxies, 
luminosity evolution is expected to have negligible effects.
Second, our first adjustment to the $z_{\rm min}$ calculation narrows the distribution 
of cloned random galaxies. Our test findings in Section~\ref{sec:comparison} will 
demonstrate that the smoothed density-corrected \vmax\ approach is adequate 
for obtaining accurate galaxy clustering determination. 

\section{Mock Galaxy Catalogs} \label{sec:mocks}

In this section, we describe the construction of mock galaxy catalogs 
for a robust test of the smoothed density-corrected \vmax\ approach 
on clustering estimation. We build two sets of mock samples, 
one with simple $k+e$ corrections and the other with complex 
$k+e$ corrections for galaxies. 

The first group of mock galaxy catalogs is created in a manner similar 
to that used in \citetalias{2020RAA....20...54Y}. For the halo catalog, we adopt the 
$\rm WMAP\_3072\_600$ cosmological $N$-body simulation from 
the CosmicGrowth simulation suite \citep{2019SCPMA..6219511J}.
This simulation starts at redshift 144 with $3072^3$ particles evolving 
in a $600~\mpch$ cube box. The simulation assumes a standard flat 
$\Lambda \rm CDM $ cosmology with $\{\Omega_m=0.268,~\Omega_b=0.045, 
~\sigma_8=0.83,~n_s=0.968\}$ and $h=H_0/(100\,\kms\mathrm{Mpc}^{-1})=0.71$, 
which are compatible with the Nine-Year Wilkinson Microwave Anisotropy Probe
(WMAP 9) observations \citep{2013ApJS..208...19H,2013ApJS..208...20B}. 
This simulation has a mass resolution of $5.54 \times 10^8~h^{-1}\rm M_{\sun}$. 
To identify halos for each output snapshot, the friends-of-friends technique is 
used with a linking length of 0.2 in units of the mean particle separation 
\citep{1985ApJ...292..371D}. The Hierarchical Bound-Tracing technique is used to find 
subhalos and their merger histories. In this study, the snapshot 
at $z=0$ is utilized to build the halo catalog, and each halo 
contains at least 50 particles. The {\it orphan} halos are also 
maintained in the catalog \footnote{In the evolution 
process, some subhalos go below the resolution limit due to the tidal stripping. 
We keep subhalos whose infall time is shorter than the merger time, and those 
subhalos do not merge into the core of the host halo and host 
the {\it orphan} galaxies.} \citep{2019ApJ...872...26Y}.

We use the subhalo abundance matching method to 
establish the connection between galaxies and subhalos.
Based on the galaxy's absolute magnitude $M^{0.1}_{\rm r}$ 
and the peak mass $M_{\rm peak}$ of subhalos, a monotonic relationship between 
the cumulative number density $n(<M^{0.1}_{\rm r})=n(>M_{\rm peak})$ 
has been constructed \citep{2006ApJ...647..201C,2014MNRAS.444..729H,
2018ARA&A..56..435W,2021MNRAS.508..175C}. 
We employ the LF of the SDSS DR7 \full\ sample of 
the New York University Value-Added catalog 
(NYU-VAGC)\footnote{$\tt lfvmax-q2.00a-1.00.dr72full1.fits$.}
\citep{2001AJ....121.2358B,2003ApJ...592..819B,2005AJ....129.2562B}, 
for which the $r-$band absolute magnitude $M^{0.1}_{\rm r}$ of galaxies 
has been $k$ and $e$corrected to $z=0.1$. 
The $M_{\rm peak}$ is the maximum mass ever attained by 
a subhalo over its entire evolutionary history. 
Once a subhalo has been matched to a galaxy, 
its position and velocity are given to the galaxy. 
By periodically rotating and stacking the mock box, we generate 60 mock galaxy 
catalogs from the parent catalog. Random sites are assigned to the observer. 
The observed redshift $z_{\rm obs}$ is determined by the galaxy's position 
and velocity relative to the observer. To obtain the apparent magnitude $m_{\rm r}$, 
the $k$ correction and $e$ correction, as described in 
Equations~\eqref{eq:mm1} and \eqref{eq:mm2}, must be provided.
Real data processing determines these values by fitting the observed galaxy flux 
to a library of synthetic spectrum models, 
which is generally inapplicable to mock galaxies and also beyond the scope of this work. 
For the sake of simplicity, we consider two simple $k+e$correction cases here. 
In the first case, no $k+e$ corrections are applied to the mock galaxies. 
In the second case, we suppose that all galaxies follow a simple $k+e$ correction model.
For the $k$correction, we take the model of \cite{2017MNRAS.470.4646S}: 
\begin{equation}\label{eq: kcor}
k^{0.1}(z)=\sum^{4}_{i=0} A_i(z-0.1)^{4-i}.
\end{equation}
\cite{2017MNRAS.470.4646S} fit the above fourth-order polynomial 
to individual GAMA galaxies, where $A_i$ is the polynomial's 
fitting coefficient \citep{2014MNRAS.445.2125M}. 
There are seven color-dependent $k(z)$ models (see the section below) and 
we adopt the $(g-r)^{0.1}_{\rm med} =0.603$ model with the following 
fitting coefficients: $A_0=-3.428$, $A_1=9.478$, $A_2=-2.703$, and $A_3=0.7646$.
For the $e$ correction, we use the SDSS model \citep{2006ApJ...648..268B} : 
\begin{equation}\label{eq:ecor}
E(z)=q_0[1+q_1(z-z_0))](z-z_0),
\end{equation}
where $z_0=0.1$ is the zero-point redshift for evolution correction, $q_0=2$ 
denotes the evolution of magnitude per redshift, $q_1= -1$ is 
the nonlinear parameter in redshift evolution. 
After applying the $k+e$ corrections to the mock galaxies, 
our final samples are constructed as follows. 
For each mock catalog in each $k+e$ correction case, we first generate 
a flux-limited sample with flux cuts at $m_{\rm r} = [15, 17]$ and 
a sky coverage of $\sim 5950~\rm deg^2$. 
The flux-limited catalog is then divided into two luminosity-dependent samples, 
designated LC1 with $M^{0.1}_{\rm r} = [-19, -22]$ and LC2 with $M^{0.1}_{\rm r} = [-20, -23]$.
Using these selection criteria, the galaxy sample's number density 
changes as a function of redshift. 
 Figure~\ref{fig:nd_mocks} in Appendix~\ref{sec:append_nd} displays 
 the average number density $\overline{n}(z)$ of the 60 samples for two luminosity cuts 
 in each $k+e$ correction case. This redshift-dependent number density typically prevents 
 us from obtaining an accurate measurement of galaxy clustering, particularly at scales 
$\le 30\mpch$ for flux-limited samples \citep{2022arXiv221102068Y}. 
In the following text, the above mock samples generated from 
the simulation of \citep{2019SCPMA..6219511J} are referred to as LC samples.

The second group of mock galaxy catalogs is built from
the light cone catalog of \cite{2017MNRAS.470.4646S} 
\footnote{http://icc.dur.ac.uk/data/}. 
It is essential to assess the radial selection model using 
a catalog of galaxies that closely resembles the observed galaxies. 
The \cite{2017MNRAS.470.4646S} catalog is constructed using 
the MXXL simulation \citep{2012MNRAS.426.2046A}, which assumes 
a $\Lambda$CDM cosmology with WMAP1 parameters 
$\{\Omega_m=0.25, ~\sigma_8=0.9,~n_s=0.968, h=0.73\}$ and operates in a 
$3h^{-1}\rm Gpc$ box. 
The mass of the particle is $6.17\times 10^9 h^{-1}\rm M_{\sun}$. 
\cite{2017MNRAS.470.4646S} created the light cone catalog by applying 
the halo occupation distribution method to link galaxies to subhalos. 
To assign colors to the galaxies, they utilize an enhanced redshift-dependent 
\cite{2009MNRAS.392.1080S} model. The galaxy $k+e$ corrections in their 
light cone catalog are more complicated than the ones we use for the LC samples.
They employ color-dependent $k$ corrections obtained from the GAMA survey for 
the $k$ corrections. In brief, they estimate the $k$ corrections for individual galaxies 
in GAMA data by fitting with equation~\eqref{eq: kcor}, and they determine 
the median $k$ correction in seven evenly spaced color bins to construct 
seven $k-$correction models. These models are 
$(g-r)^{0.1}_{\rm med} = $ 0.131,0.298,0.443,0.603,0.785,0.933, and 1.067 
with different polynomial coefficients. The $k(z,color)$ is then interpolated 
for the light cone catalog using seven median color $(g-r)^{0.1}_{\rm med}$ 
models based on the galaxy's color and redshift \footnote{
For details see Setion4.3 in \cite{2017MNRAS.470.4646S}}. 
For the LF evolution, they employed the evolving Schechter function 
derived from GAMA data. In the low-redshift region $z \lesssim 0.13$, 
the LF of their catalog coincides with the LF of \cite{2003ApJ...592..819B}, 
which we employ for the LC samples, and in the median redshift region, 
the LF evolves to the GAMA LF. The luminosity(color)-dependent galaxy 
clusterings in \cite{2017MNRAS.470.4646S} catalog are generally consistent with 
the SDSS DR7 results measured by \cite{2011ApJ...736...59Z} at low redshift, 
as well as the GAMA results measured by \cite{2015MNRAS.454.2120F} 
at the median redshift. Therefore, this catalog is suitable for testing different 
radial selection models for property-dependent clustering measurement. 
We construct 10 flux-limited samples from the full-sky light cone catalog 
by rotating the sky, using the galaxy selection criteria 
($m_{\rm r} = [15, 17]$) and sky coverage ($\sim 5950~\rm deg^2$). 
Two luminosity-dependent galaxies, LS1 ($M^{0.1}_{\rm r} = [-19, -22]$) 
and LS2 ($M^{0.1}_{\rm r} = [-20, -23]$), are generated from each flux-limited sample, 
much as we did for the LC samples. As our sample selection resembles 
the SDSS DR7 data, we further divide the luminosity-dependent sample into 
a blue subsample and a red subsample using the color-cut equation 
$(g-r)^{0.1}_{\rm cut}=0.21-0.03M^{0.1}_{\rm r}$ of \cite{2011ApJ...736...59Z}.
In the rest of this study, we refer to the mock galaxy samples 
built from the \cite{2017MNRAS.470.4646S} catalog as LS samples.

In summary, we generate two sets of mock samples from 
two simulations using the same selection criterion for galaxies. 
For the LC samples, flux-limited samples are constructed from 
60 mocks with two absolute magnitude cuts. 
Two cases are considered for $k+e$ corrections: (1) there are no $k+e$ corrections; 
(2) all galaxies are assumed to follow a simple $k+e$ correction model. 
Ten LS samples are created in the same manner as the LC samples, 
but using a public light cone catalog. 
The LS samples, however, feature a color-dependent $k-$correction and 
a complex $e$ correction that are unknown to us. In order to examine 
the color-dependent clustering, the luminosity-dependent LS data are split 
into blue and red subsamples. We emphasize that neither the LC samples 
nor the LS samples are subjected to any deliberate impact (e.g., fiber collision) 
in order to decrease unknown systematic uncertainty in our later tests. 
In addition, when calculating the comoving distance from redshift, 
we employ the cosmological model of the simulation from which the samples 
are constructed, separately.


\section{Testing the Smoothed Density-corrected \vmax\ Method with the 2PCFs} \label{sec:tests}

In this section, we describe the construction of a random galaxy catalog, 
focusing on the radial distribution of random galaxies derived from 
various radial selection models. Following that, we compare the correlation functions 
generated by the random catalogs used in these models.

\subsection{Construction of the Random Catalogs} \label{sec:randoms}

The random catalogs are constructed as follows. 
For the angular distribution, we first generate a large number of 
random points that are uniformly dispersed across the surface of 
a unit sphere. For each mock sample and subsample, we extract 
a collection of points with the same sky coverage as the corresponding 
sample and subsample. We consider the positions of these points to be the angular distribution 
(R.A., decl.) of the random galaxies, with no angular selection effect or survey 
masks imposed. For the redshifts of random galaxies, 
the following radial selection models are used in our tests:

\begin{enumerate}
      \item $\mathbf{n_{\rm \textbf{true}}}$ method, which generates the redshift 
      distribution for random galaxies using the true galaxy number density 
      $n(z)_{\rm true}$ taken from the LF of the parent mock catalog.
      \item $\mathbf{V^{\rm \textbf{SDC}}_{\rm \textbf{max}}}$ method, in which
      redshifts for the random catalog are generated using 
      the smoothed density-corrected \vmax\ method.
      \item $\mathbf{V^{\rm \textbf{DC}}_{\rm \textbf{max}}}$ method, in which 
      the density-corrected \vmax\ method of \cite{2011MNRAS.416..739C} is utilized, 
      but without the smoothing procedure.
      \item $\mathbf{V_{\rm \textbf{max}}}$ method, where the normal \vmax\ method is adopted. 
      \item \textbf{Shuffled} method, which applies the redshift shuffled method. In this method, 
      galaxy redshifts of the sample are randomly assigned to the random galaxies.
\end{enumerate}

\begin{figure*}
\begin{center}
\centering
  \epsscale{.8}
  \plotone{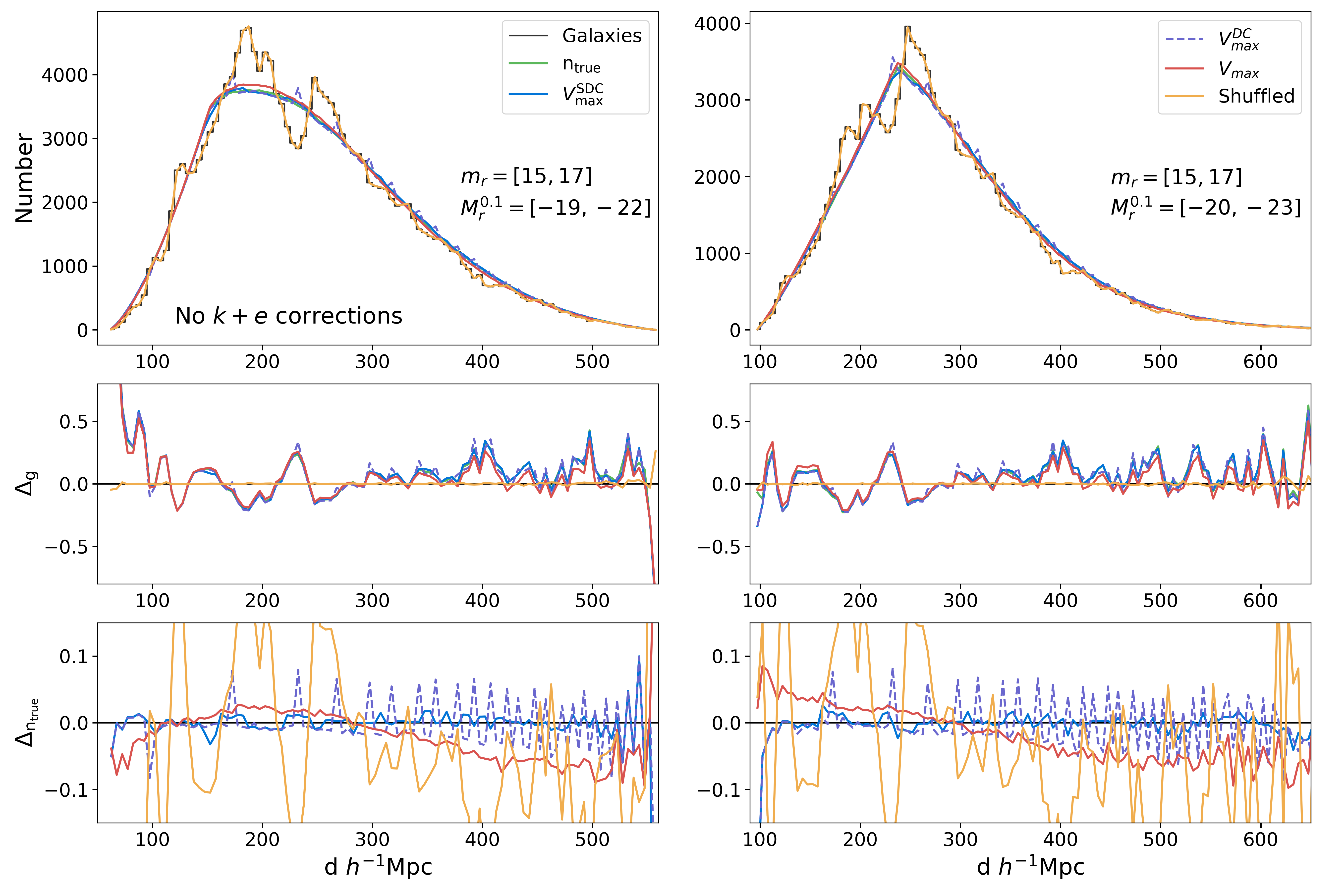}
  \caption{In the case of no $k+e$ correction, a comparison of the radial 
  distributions of one LC sample and its corresponding random catalogs.
  The bin size is $\Delta d=5~\mpch$. The LC samples have 
  a flux cut at $m_{\rm r}=[15,17]$ and two luminosity cuts at 
  $M^{0.1}_{\rm r}=[-19,-22]$ (left panels) and $M^{0.1}_{\rm r}=[-20,-23]$ 
  (right panels). The black histogram denotes galaxy distribution. 
  Random catalogs generated by the $n(z)_{\rm true}$, 
  \sdcvmax, \dcvmax, \vmax, and shuffled methods are represented by the green, blue, 
  purple-dashed, red, and yellow lines, respectively. 
  The second row of panels displays the number bias $\Delta_{\rm g}$ 
  in each bin of the random catalogs compared to the galaxies, 
  calculated as $\Delta_{\rm g} = (n_{\rm r}-n_{\rm g})/n_{\rm g}$. 
  The third row of panels displays the number bias of random catalogs 
  compared to $n(z)_{\rm true}$, which is defined as 
  $\Delta_{\rm n_{\rm true}}=(n_{\rm r}-n_{\rm r, true})/n_{\rm r, true}$. }
\label{fig:histLC1}
\end{center}
\end{figure*}

For LC samples, it is simple to incorporate the $k+e$ corrections 
into the redshift generation process. 
Enabling the validation of the capacity of different radial selection 
models to restore the true radial selection function $n(z)_{\rm true}$. 
Figure~\ref{fig:histLC1} shows a comparison between the radial distributions of a single 
LC sample and random catalogs generated by the aforementioned radial selection methods 
in the case of no $k+e$ corrections. In the left and right panels, the comparisons 
for LC1 and LC2 samples are presented, respectively. 
The second row of panels displays the deviation of random galaxy 
number relative to the galaxy number in each comoving distance bin, 
which is defined as $\Delta_{\rm g} = (n_{\rm r}-n_{\rm g})/n_{\rm g}$. 
The third row of panels displays the deviation of the random galaxy number 
of the other four techniques from the number of the $n_{\rm true}$ approach, 
defined as $\Delta_{\rm n_{\rm true}}=(n_{\rm r}-n_{\rm r, true})/n_{\rm r, true}$. 
The black histograms in the top row of panels denote the distribution of galaxies 
in the flux-limited samples. The radial distribution of random catalogs created by 
the $n_{\rm true}$ method is represented by green lines, which indicate 
the distribution arising from the genuine selection function. 
The purple-dashed line indicates the distribution produced from the \dcvmax\ approach. 
We see small fluctuations in the radial distribution, which are notably clear 
in the bottom row of panels. These noisy fluctuations have been reduced 
by the smoothing process in the \sdcvmax\ approach; as indicated by 
the blue solid lines, the smoothed radial distribution is in excellent 
agreement with the distribution predicted by the $n_{\rm true}$ method.

\begin{figure*}
\begin{center}
\centering
  \epsscale{.8}
  \plotone{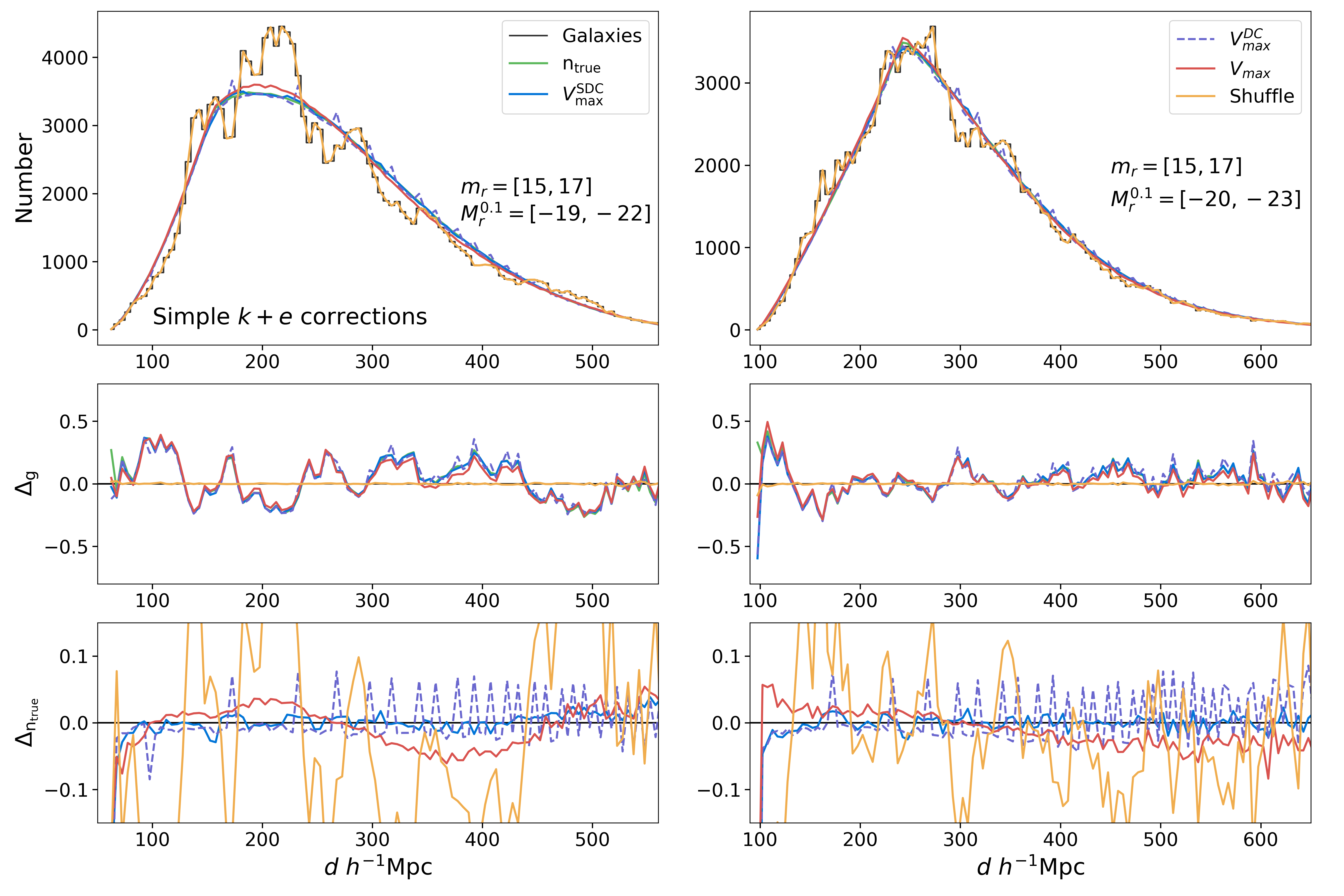}
  \caption{The same as Figure~\ref{fig:histLC1} but for the simple $k+e$ correction case of the LC samples. }
\label{fig:histLC2}
\end{center}
\end{figure*}

The radial distributions from the \vmax\ and the shuffled methods 
are represented by red and yellow lines, respectively. 
As shown in the bottom panels, $\Delta_{\rm n_{\rm true}}$ of 
the \vmax\ approach exhibits a systematic bias in both 
luminosity-dependent LC samples as a result of the influence of 
large-scale structures in galaxy radial distribution. 
The \vmax\ approach creates an excess of random galaxies near 
these structures; hence, the number of random galaxies 
in the high-redshift tail has been decreased. 
Figure~\ref{fig:histLC2} shows the same comparison as 
Figure~\ref{fig:histLC1} for LC samples with the simple $k+e$ corrections.
The deviations of different approaches from the $n_{\rm true}$ method 
shown in the bottom panels are similar to those in Figure~\ref{fig:histLC1}.

\begin{figure*}
\begin{center}
\centering
  \epsscale{1.}
  \plotone{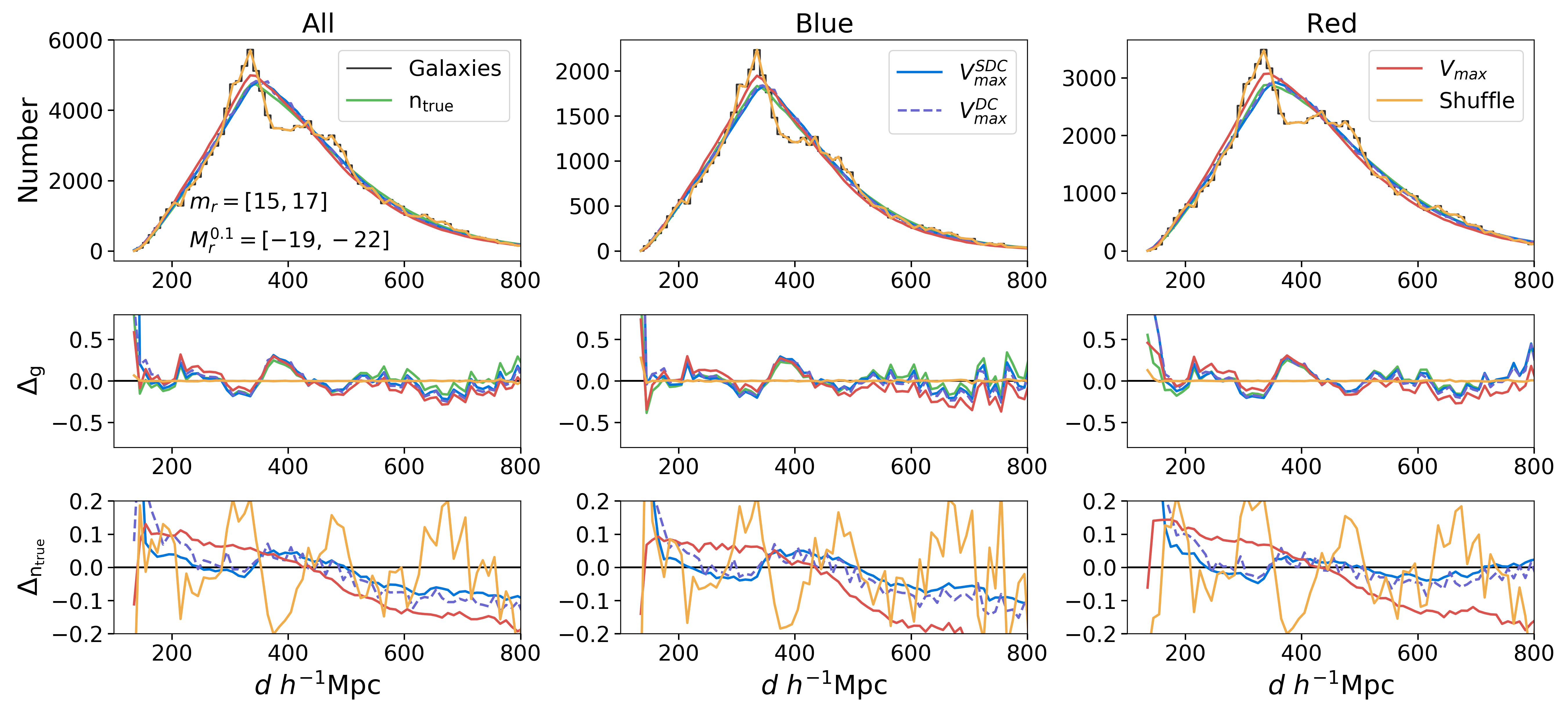}
  \caption{The same as Figure~\ref{fig:histLC1} but for the LS1 samples. }
\label{fig:histLS1}
\end{center}
\end{figure*}

Figure~\ref{fig:histLS1} shows a comparison for the LS samples, employing 
the same color-coded lines as Figure~\ref{fig:histLC1}. The left panels compare 
an LS1 sample, whereas the middle and right panels compare its blue and red 
subsamples, respectively. 
For the $n_{\rm true}$ method, the radial selection function derived from 
the LF of the light cone catalog is applied. The $k+e$ corrections are appropriately 
incorporated into the redshift generation process for the $n_{\rm true}$ and \vmax\ methods.
For the \sdcvmax\ and \dcvmax\ methods, 
the same $k-$correction models that \cite{2017MNRAS.470.4646S} performed for 
their light cone database are employed, which interpolate the $k$ correction 
from seven models based on the color and redshift of individual galaxies. 
The $e$ correction is also properly applied to the LS samples and 
their color-dependent subsamples by using the evolutionary property 
of the light cone catalog.
The results of the comparison are generally consistent with those of the LC samples.
The redshifts generated by the \vmax\ technique are substantially influenced by 
the sample's structures; the bias in $\Delta_{\rm n_{\rm true}}$ is greater than that 
of the LC samples, which reaches 20\% on the high-redshift tail (red solid lines). 
The redshifts from the \sdcvmax\ approach successfully mitigate this impact, 
resulting in a relatively small deviation in $\Delta_{\rm n_{\rm true}}$ (blue solid lines). 
For both the LC and LS samples, the redshifts of random catalogs obtained by 
the shuffled approach replicate the radial distribution of galaxies (yellow solid lines); hence, 
the structures are also cloned. In the following section, we will examine how galaxy clustering 
measurements are affected by the deviations in these radial distributions that differ 
from the expected distribution produced by the $n_{\rm true}$ model.


\subsection{Comparison of the Correlation Functions} \label{sec:comparison}

This section introduces the 2PCF estimator that we employ to measure 
galaxy clustering. Then, we provide a comparison of the projected 2PCFs and 
the redshift-space 2PCFs determined from random catalogs generated 
by the aforementioned radial selection methods.

\subsubsection{Estimator} \label{sec:estimator}

We measure the 2PCF in the same way as \citetalias{2020RAA....20...54Y}. 
First, we define the redshift separation vector 
$\bm{s}$ and the line-of-sight vector $\bm{l}$ as 
$\bm{s} \equiv \bm{\upsilon}_1-\bm{\upsilon}_2 $ and
$\bm{l}\equiv (\bm{\upsilon}_1+\bm{\upsilon}_2)/2$, 
where $\bm{\upsilon}_1$ and $\bm{\upsilon}_2$ are redshift-space 
position vectors of a pair of galaxies \citep{1992ApJ...385L...5H, 1994MNRAS.266...50F}.
Separations that are parallel ($\pi$) and
perpendicular ($r_p$) to the line-of-sight direction are derived as
\begin{equation}
 \pi \equiv \frac{\bm{s}\cdot \bm{l}}{|\bm{l}|}, ~~~~~~r^2_p \equiv \bm{s}\cdot \bm{s}-\pi^2.
\end{equation}
 We construct a grid of $\pi$ and $r_p$ by taking $1~\mpch$ as the bin size for $\pi$ from 
 0 up to $\pi_{\rm max}=40~\mpch$ linearly, and a bin size of $0.2$ for $r_p$ is adopted 
logarithmically in the range of [$0.01$, $40$] $\mpch$. 
The estimator of \cite{1993ApJ...412...64L} is used to calculate the 2D correlation function as
\begin{equation}
 \xi(r_p,\pi)  = \frac{DD-2DR+RR}{RR},
\end{equation}
where $DD$, $DR$, and $RR$ are the numbers of data-data, data-random,
and random-random pairs. Given $s^2=| \bm{s}|^2=r^2_p +\pi^2$, we derive 
the redshift-space correlation function $ \xi(s)$.
By integrating $\xi(r_p,\pi)$ along the line-of-sight direction, 
we estimate the projected 2PCF \citep{1983ApJ...267..465D} by 
\begin{equation}
w_p(r_p)\equiv 2\int^{\infty}_0 \xi(r_p,\pi)~d\pi \approx 2\int^{\pi_{max}=40}_0 \xi(r_p,\pi)~d\pi.
\end{equation}
We employ the public code $\tt{CORRFUNC}$ \citep{10.1007/978-981-13-7729-7_1} 
for pair counting in this work. To reduce the shot noise on small-scale clustering, 
we use 50 times the number of galaxies in the random catalogs for random galaxies.

\subsubsection{Comparison of Projected 2PCFs} \label{sec:compwps}

\begin{figure*}
\begin{center}
\centering
  \epsscale{1.}
  \plotone{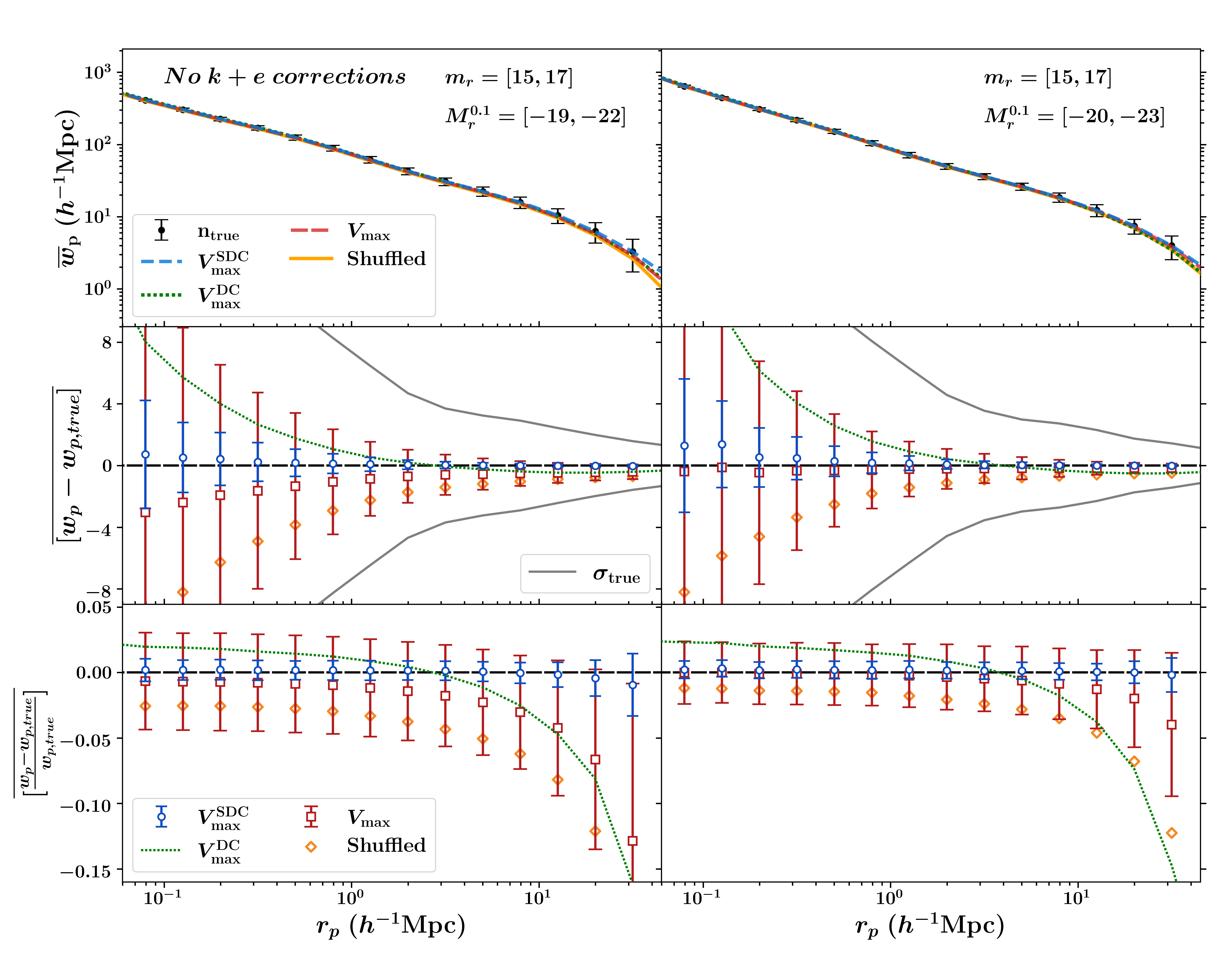}
  \caption{Top panels: The average projected correlation functions 
  $\overline{w}_p$ for LC1 (left panel) and LC2 (right panel) samples 
  in the case of no $k+e$ corrections. 
  The LC1 samples have a flux-cut at $m_r=[15,17]$ and 
  a luminosity cut at $M^{0.1}_r=[-19, -22]$. 
  LC2 samples have the same flux-cut as LC1 samples 
  but a brighter luminosity cut at $M^{0.1}_r=[-20, -23]$. 
  The solid black points with error bars represent the $\overline{w}_{p,true}$ 
  and $1\sigma$ dispersion across 60 of the LS samples utilizing 
  random catalogs generated by the $n_{\rm true}$ approach. 
  $\overline{w}_p$ of the \sdcvmax, \dcvmax, 
  \vmax, and shuffled methods are shown by 
  the blue-dashed lines, the green-dotted lines, 
  the red long-dashed lines, and the orange lines, respectively. 
  Middle panels: the average deviations from $w_{p,true}$ 
  for various techniques of assigning redshifts to random catalogs, 
  as determined by the $w_p$ of the 60 LC samples. 
  The blue open rolls with error bars represent 
  the mean offset and $1\sigma$ deviations of $w_p$ 
  for the \sdcvmax\ technique. The results of 
  the \vmax\ technique are displayed as open red squares 
  with error bars. The mean offsets computed from $w_p$ 
  for the \dcvmax\ and shuffled methods are shown by 
  green-dashed lines and yellow open diamonds, accordingly. 
  The gray lines represent the $1\sigma$ dispersion 
  of $w_{p,true}$ among the 60 LC samples. 
  The horizontal-dashed black lines indicate the zero offset. 
  Bottom panels: The average bias of $w_p$ 
  relative to $w_{p,true}$ for four radial selection models, 
  defined as $\overline{[(w_p-w_{p,true})/w_{p,true}]}$. 
  The color-coded lines and symbols are identical to those 
  in the middle panels.}
\label{fig:wp_noke}
\end{center}
\end{figure*}

\begin{figure*}
\begin{center}
\centering
  \epsscale{1.}
  \plotone{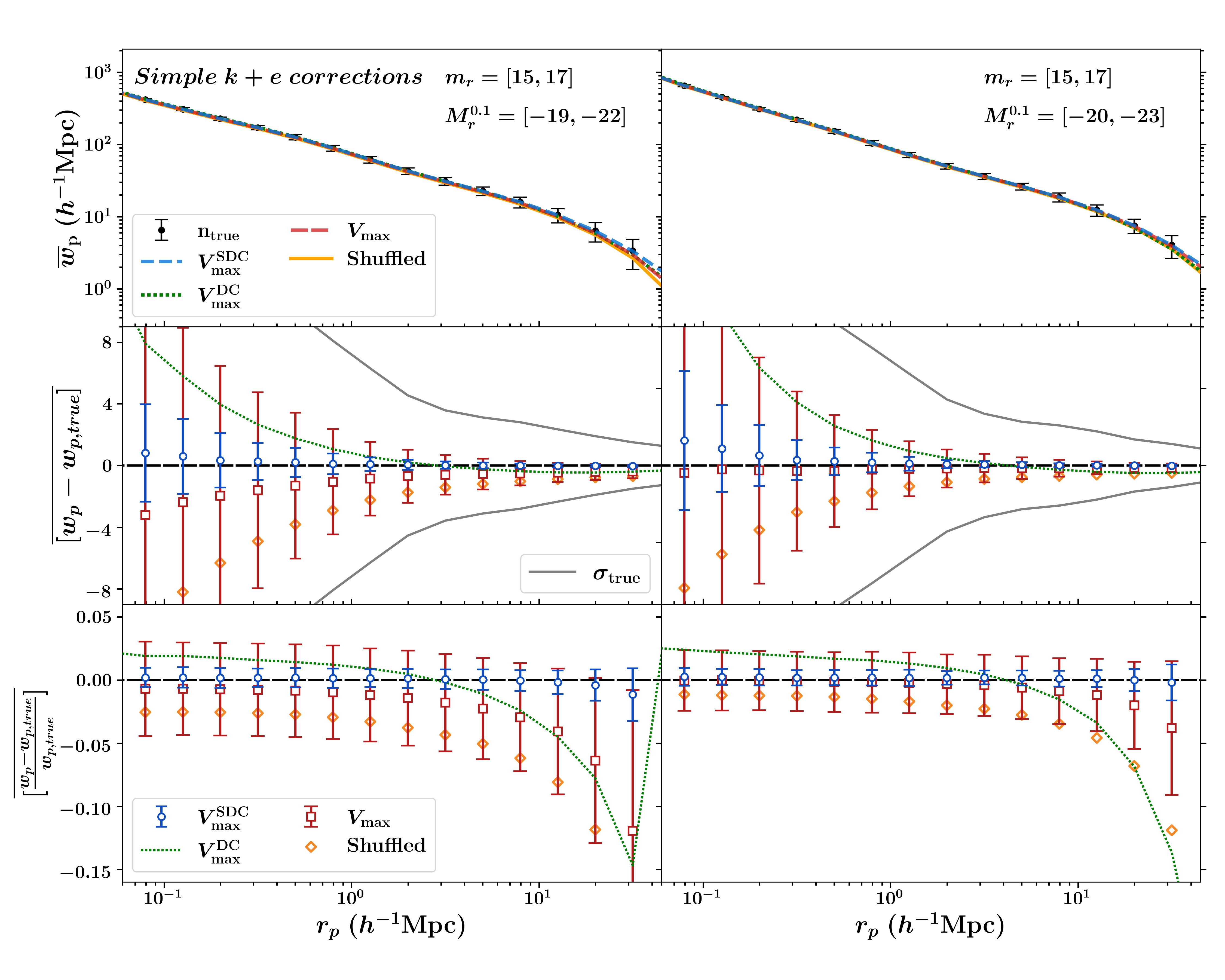}
  \caption{The same as Figure~\ref{fig:wp_noke} but for the LC samples with simple $k+e$ corrections.}
\label{fig:wp_kpluse}
\end{center}
\end{figure*}

\begin{figure*}
\begin{center}
\centering
  \epsscale{1.}
  \plotone{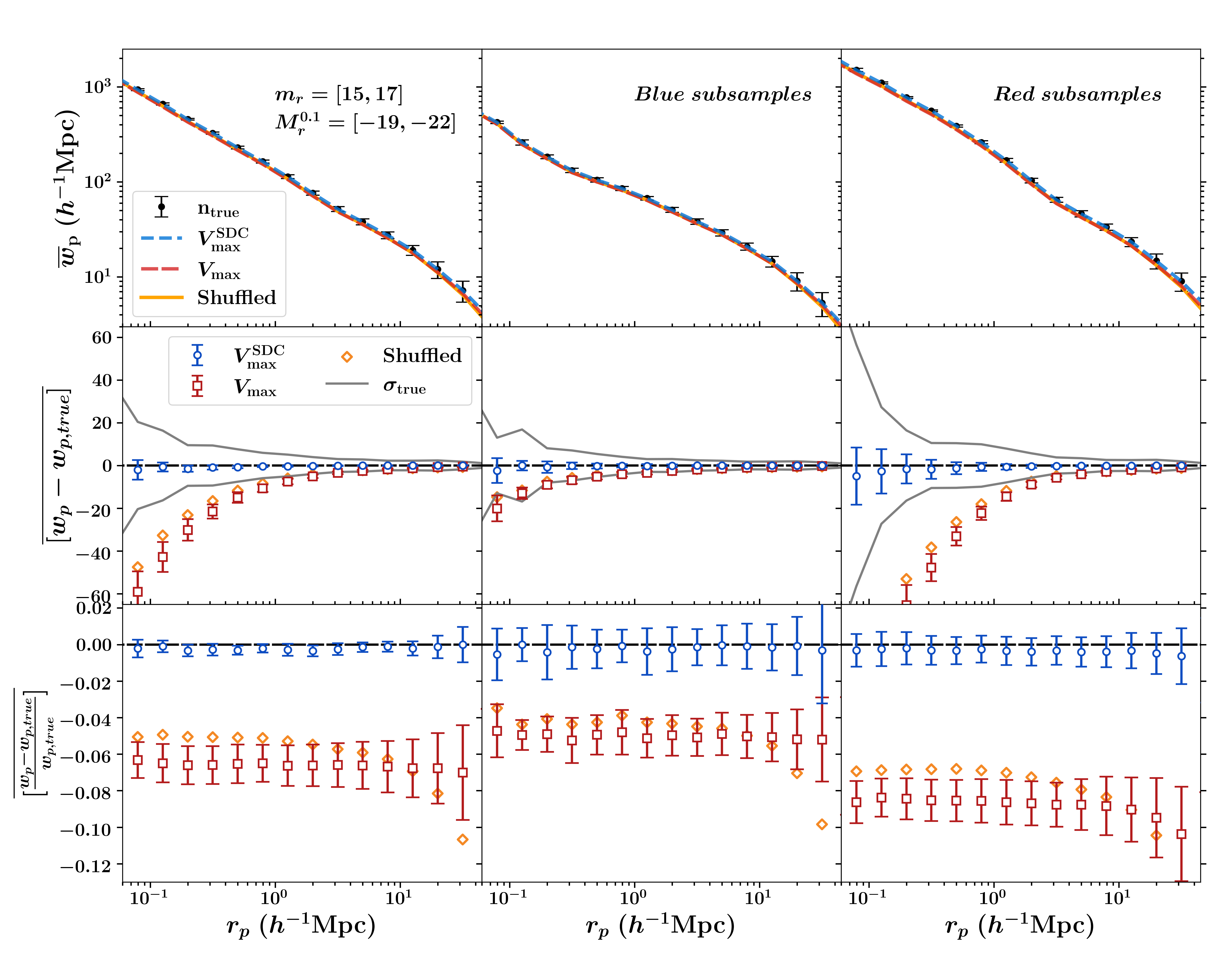}
  \caption{Similar to Figure~\ref{fig:wp_noke}: comparison of $w_p$ 
  for the LS1 samples (left panels) and their blue (middle panels) 
  and red (right panels) subsamples. 
  The color-coded lines and symbols are identical to those in 
  Figure~\ref{fig:wp_noke}, excluding the result of 
  the \dcvmax\ technique.}
\label{fig:wp_LS1}
\end{center}
\end{figure*}

\begin{figure*}
\begin{center}
\centering
  \epsscale{1.}
  \plotone{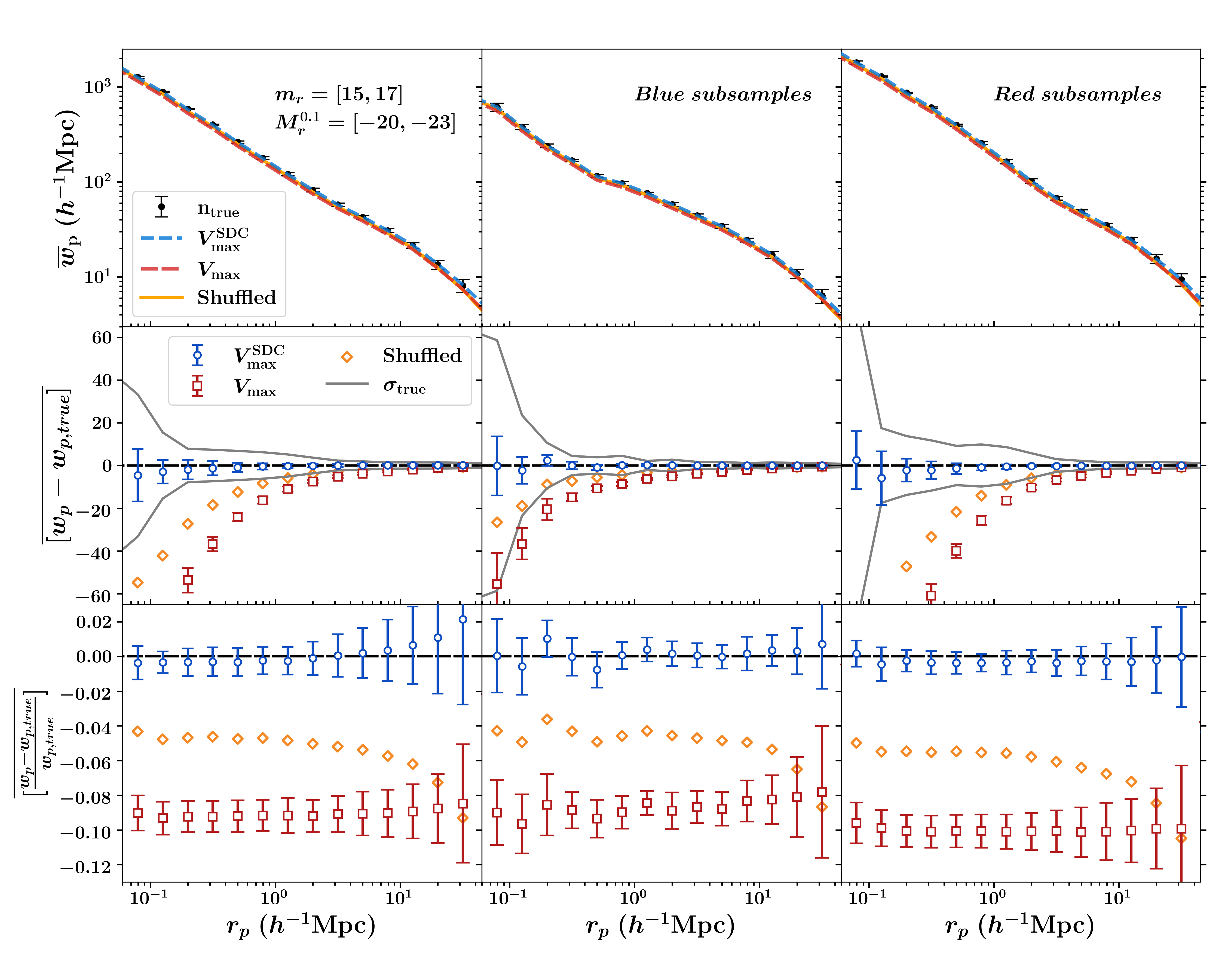}
  \caption{The same as Figure~\ref{fig:wp_LS1} but for the LS2 samples (left panels) 
  and their blue (middle panels) and red (right panel) subsamples.}
\label{fig:wp_LS2}
\end{center}
\end{figure*}

 The projected 2PCFs for the LC samples without and with simple $k+e$ corrections 
 are compared in Figure~\ref{fig:wp_noke} and Figure~\ref{fig:wp_kpluse}, respectively. 
 We compare the average projected 2PCF estimated using random catalogs produced by the radial selection 
 models outlined in Section~\ref{sec:randoms}. In the left and right panels for the LC1 and LC2 samples, 
 respectively, the estimated mean $\overline{w}_p$ of the 60 mock samples are displayed.
 In the top panels, $\overline{w}_{p,true}$ computed using the 
 random catalog from the $n_{true}$ model is represented 
 by solid black points with errors representing the $1\sigma$ dispersion 
 across individual $w_{p,true}$s of samples. The blue-dashed, green-dotted, 
 red long-dashed, and orange lines represent $\overline{w}_{p}$s estimated 
 from random catalogs of the \sdcvmax, \dcvmax, \vmax, 
 and shuffled methods, respectively. The average offsets $\overline{[w_p-w_{p,true}]}$ 
 from $w_{p,true}$ for the models are shown in the middle row of panels, 
 which are defined as $\overline{[w_p-w_{p,true}]}=\frac{1}{60}\sum^{60}_{i=1}{(w^i_p-w^i_{p,true})}$, 
 where $w^i_p$ is the projected 2PCF measured for the $i$th LC sample.
 The offsets increase when the scale drops below $1\mpch$ for both the \dcvmax\ method 
 (green-dotted lines) and the shuffled method (orange diamonds). When using the random catalogs 
 of the \sdcvmax\ technique to measure $w_p$, the little positive offsets in the blue open rolls 
 with error bars indicate a slight overestimation on scale $r_p  \lesssim 0.4\mpch$. 
 On a small scale, there are apparent offsets for the \vmax\ approach for LC1 samples 
 in both $k+e$ correction cases, as can be seen by the open red squares with error bars. 
 For the LC2 samples, there are extremely modest systematic offsets for the \vmax\ technique 
 across all of the scales tested, and these offsets are smaller than those for the \sdcvmax\ method. 
 Compared to the $1\sigma_{true}$ (gray solid lines) among 60 $w_{p,true}$s, 
 the \sdcvmax\ and \vmax\ methods' offsets are essentially insignificant.

In the bottom panels of Figure~\ref{fig:wp_noke} and Figure~\ref{fig:wp_kpluse}, 
we display the average deviation from $w_{p,true}$ 
for each model, using the same color-coded symbols and lines as the middle panels. 
The mean deviation $\overline{[(w_p-w_{p,true})/w_{p,true}]}$ is calculated from the 60 mock 
samples in the same manner as $\overline{[(w_p-w_{p,true})]}$. Clearly, the $w_p$ 
derived using random catalogs from the \sdcvmax\ approach provide a mostly unbiased 
estimate of the genuine projected 2PCFs for both the LC1 and LC2 samples in both the 
no $k+e$ correction case (Figure~\ref{fig:wp_noke}) and the simple $k+e$ correction case 
(Figure~\ref{fig:wp_kpluse}). The $1\sigma$ deviations among the 60 samples 
for the \sdcvmax\ approach (blue error bars) are significantly smaller than those 
for the \vmax\ method (red error bars). For the LC1 samples in both $k+e$ correction cases, 
the \vmax\ approach underestimates $w_p$ by less than 1\%, and this bias worsens as the scale grows. 
At $r_p\sim 30\mpch$, the bias reaches 13\% with a substantial variance 
\footnote{This bias is marginally less than the 20\% bias found for the \vmax\ approach 
 by \citetalias{2020RAA....20...54Y}. This may be owing to the increase in the number of galaxies 
 in the samples, as the LC samples cover twice as much sky as the flux-limited samples 
 in \citetalias{2020RAA....20...54Y}.}.
For the LC2 samples, the measurement accuracies for both the \sdcvmax\ and \vmax\ methods 
are equivalent at scale $r_p\lesssim 4\mpch$ for both methods. On a larger scale, the deviation of 
the \vmax\ method grows to 4\%, but remains within the margin of error. 
These discrepancies in $w_p$ from $w_{p,true}$ for the \vmax\ model are mostly 
attributable to density fluctuations in galaxy samples. The $w_p$ measured 
using random catalogs from the \dcvmax\ approach are overestimated at scale 
$r_p \lesssim 2\mpch$ and underestimated at larger scales for both the LC1 and LC2 samples as shown 
in the bottom panels (green-dashed lines) of Figure~\ref{fig:wp_noke} and Figure~\ref{fig:wp_kpluse}. 
As can be seen in Figure~\ref{fig:histLC1} and Figure~\ref{fig:histLC2}, this tendency of deviation is 
the result of small fluctuations in the radial distribution of the random catalog generated 
by the \dcvmax\ model. In essence, the fluctuations increase the number of RR pairs 
at the fluctuation scale, resulting in an underestimation of $w_p$. 
Due to the integral constraint effect, a small-scale overestimation of $w_p$ is unavoidable. 
After smoothing out the fluctuations, the \sdcvmax\ approach yields estimates that are almost 
unbiased of $w_{p,true}$. The results of the shuffled technique are consistent 
with \citetalias{2020RAA....20...54Y}, which shows that an underestimation of 
$w_p$ grows as the scale increases.

Due to the severe deviations of $w_p$ for the \dcvmax\ model in the tests 
using the LC samples, the following comparison for the LS samples will focus on 
testing for the \sdcvmax, \vmax, and shuffled methods.
Figure~\ref{fig:wp_LS1} and Figure~\ref{fig:wp_LS2} display the results of the comparison 
for the LS samples with the two luminosity cuts, respectively. 
The left, middle, and right panels, respectively, present $w_p$ comparisons 
for luminosity-dependent samples and their blue and red subsamples. 
From 10 mock galaxy samples, the mean $\overline{w}_p$, 
$\overline{[w_p-w_{p,true}]}$, and $\overline{[(w_p-w_{p,true})/w_{p,true}]}$ 
are calculated (from the top to bottom panels). The $n_{true}$, 
\sdcvmax, \vmax, and shuffled methods 
all utilize the same color-coded lines and symbols as those used for figures showing the LC samples. 
For the LS1 samples in Figure~\ref{fig:wp_LS1}, the \sdcvmax\ model 
produces tiny $w_p$ offsets from $w_{p,true}$, which are consistent with 
the findings for the LC samples. Significant offsets are seen for the \vmax\ and 
shuffled methods, notably for the LS1 samples and their blue subsamples, 
where the offsets are more than a $1\sigma$ dispersion of $w_{p,true}$ 
at $r_p \lesssim 3\mpch$ scale. 
The average deviations displayed in the bottom panels clearly 
demonstrate the superiority of the \sdcvmax\ approach over the \vmax\ 
and shuffled method when measuring projected 2PCFs. $\sim 0.5\%$ deviations 
are detected for both LS1 samples and their color-dependent subsamples, 
which is essentially within the $1\sigma$ error margin. For the \vmax\ approach, 
$\overline{[(w_p-w_{p,true})/w_{p,true}]}$s deviate by 6\%, 5\%, and 9\% for the LS1 samples, 
blue subsamples, and red subsamples, respectively, which are considerably larger than 
$1\sigma$ errors. At $r_p \lesssim 10 \mpch$, the mean deviations for 
the shuffled approach are marginally better than those for the \vmax\ method, 
but worsen as the scale increases, which is consistent with the test results 
for the LC samples.

Figure~\ref{fig:wp_LS2} presents a comparison of the $w_p$ for the LS2 samples. 
The offsets from $w_{p,true}$ for the \sdcvmax\ technique are roughly comparable with 
the LS1 sample results. The $w_p$ measured using random catalogs from the \vmax\ approach 
exhibit large offsets from $w_{p,true}$ that are worse than the offsets for the shuffled method 
on small scales, particularly for the LS2 samples (left middle panel) and red subsamples 
(right middle panel). 
In the bottom panels of Figure~\ref{fig:wp_LS2}, the accuracy of measurement 
for three models is shown clearly. At scale $r_p < 1\mpch$, there is a $\sim 0.5\%$ 
underestimate for the LS2 samples (bottom left panel). 
At a larger scale, this deviation becomes an overestimation, reaching 2\% at 
$r_p \sim 30 \mpch$ while being within the margin of error. 
The mean deviations for the blue and red subsamples are well constrained within 1\%. 
The results of the \vmax\ approach exhibit larger mean deviations than the LS1 samples, 
which are even worse than the results of the shuffled method. 
The deviations for the LS2 samples, blue subsamples, and red subsamples 
are roughly 9\%, 8\%, and 10\%, respectively. The $w_p$ determined for the red subsamples 
exhibit more severe departures from $w_{p,true}$ for the \vmax\ technique for 
both the LS1 and LS2 samples, demonstrating density fluctuations have a greater impact 
on clustering determination for red galaxies.

To better quantify the measurement accuracy of projected 2PCF for various radial selection models, 
we calculate the $\chi^2$ between $w_p$ and $w_{p,true}$ for the \sdcvmax, 
\vmax, and shuffled methods, respectively, as shown in Table~\ref{tab:chi2_wp}. 
$\chi2$ is computed as follows: 
\begin{equation}\label{eq:chi2}
\chi^2=\sum^{N}_{i=0} \frac{(w^i_p-w_{p,true})^2}{\sigma^2_{true}}.
\end{equation}
The number of mock samples $N$ is 60 for LC samples and 10 for the LS samples.
For the LC samples, with the exception of the LC2 samples with simple $k+e$ corrections 
for which $\chi^2$ of the \sdcvmax\ and \vmax\ methods are essentially equal, 
the $w_p$ of the \sdcvmax\ method exhibit the least $\chi^2$ from $w_{p,true}$ 
when compared to the other two models. 
For all LS samples and their blue and red subsamples, the \sdcvmax\ approach also 
yields the least $\chi^2$ among the three methods. The $\chi^2$ values for the LS samples 
are greater than those for the LC samples for all three models. 
This may probably be due to the fact that the LS samples built from a light cone catalog 
contain more complicated $k+e$ corrections than LC samples. 
On the basis of the preceding figures and $\chi^2$ tests, 
we demonstrate that the $w_p$ measured using the random catalogs 
generated by the \sdcvmax\ approach result in the least deviation from 
$w_{p,true}$ for both flux-limited samples and their color-dependent subsamples. 
In Section~\ref{sec:disc}, we discuss the performance 
of the radial selection models for the LC and LS samples.

\begin{table}[h!] 
\caption{$\chi^2$ of the Projected 2PCFs for the Mock Samples }\label{tab:chi2_wp}
\centering
\begin{tabular}{l c c  c}
\hline \hline
 Samples&  & $\chi^2$  & \\ [1ex]
\hline
 &  \sdcvmax\  &  \vmax\  & Shuffled \\ [1ex]
\hline
LC1($no~k+e$) & 1.364 &  6.264&107.225  \\
LC2($no~k+e$) & 1.460 &  4.254& 62.329 \\
LC1($simple~k+e$) & 3.531  & 6.351& 108.770 \\
LC2($simple~k+e$) & 2.757 & 2.667 &106.466 \\
\hline
LS1 & 1.893 &  1618.495 & 977.362 \\
LS1 (blue)&33.013  &  161.187 & 124.543 \\
LS1 (red)& 19.525 &   2769.991 & 1988.678 \\
LS2 & 45.168 &  3416.047& 857.843 \\
LS2 (blue)& 63.572 &   925.400 & 240.416 \\
LS2 (red)& 71.431 &  5054.464 & 1562.508 \\
\hline
\end{tabular}

\end{table}

\subsubsection{Comparison of the Redshift-space 2PCFs} \label{sec:compRSD}

\begin{figure*}
\begin{center}
\centering
  \epsscale{.8}
  \plotone{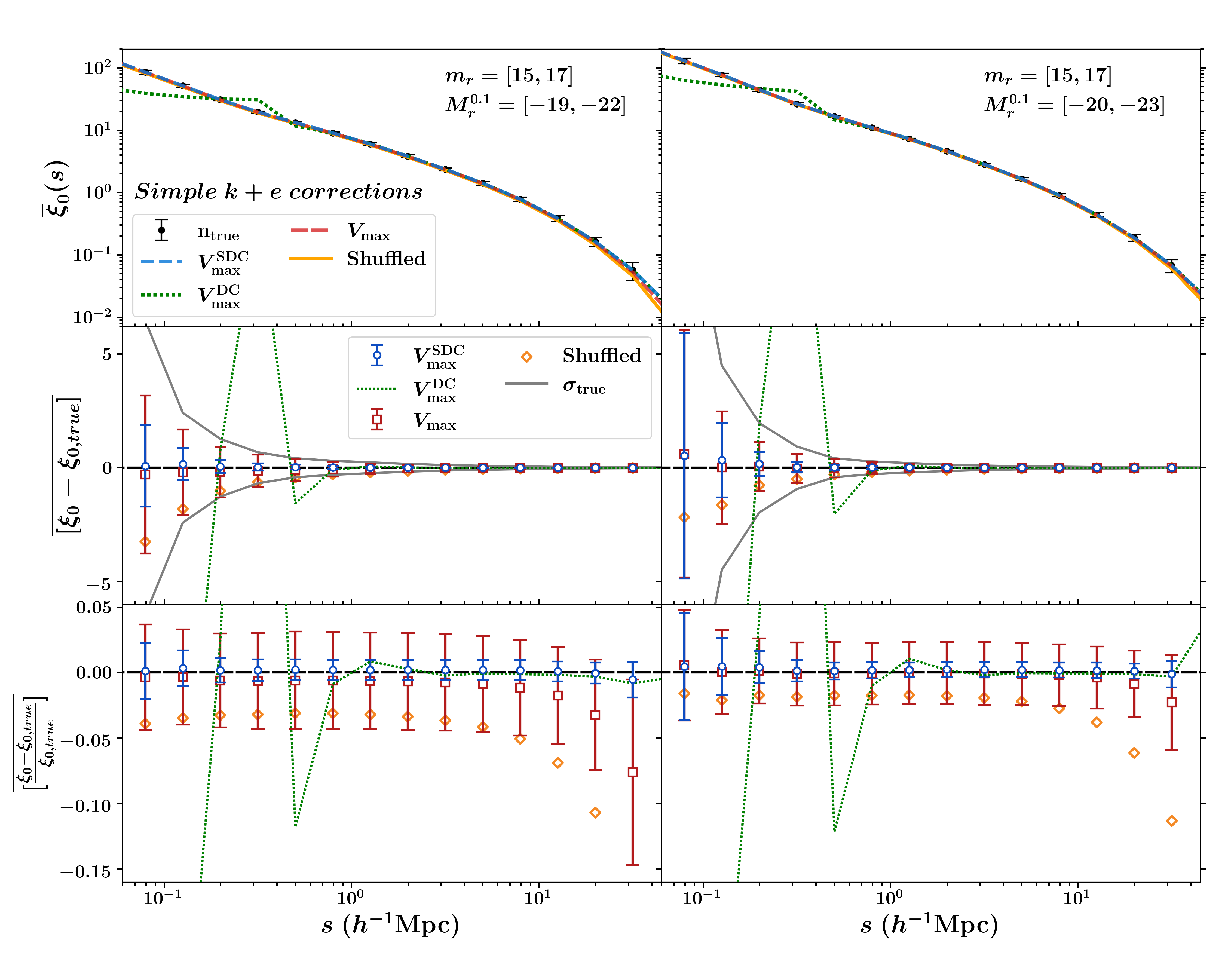}
  \caption{Similar to Figure~\ref{fig:wp_noke}, with a comparison of the $\xi_0$ 
  for the redshift-space 2PCFs of the LC1 samples (left panels) and LC2 samples 
  (right panels) with simple $k+e$ corrections.}
\label{fig:xi0_kpluse}
\end{center}
\end{figure*}

\begin{figure*}
\begin{center}
\centering
  \epsscale{1.}
  \plotone{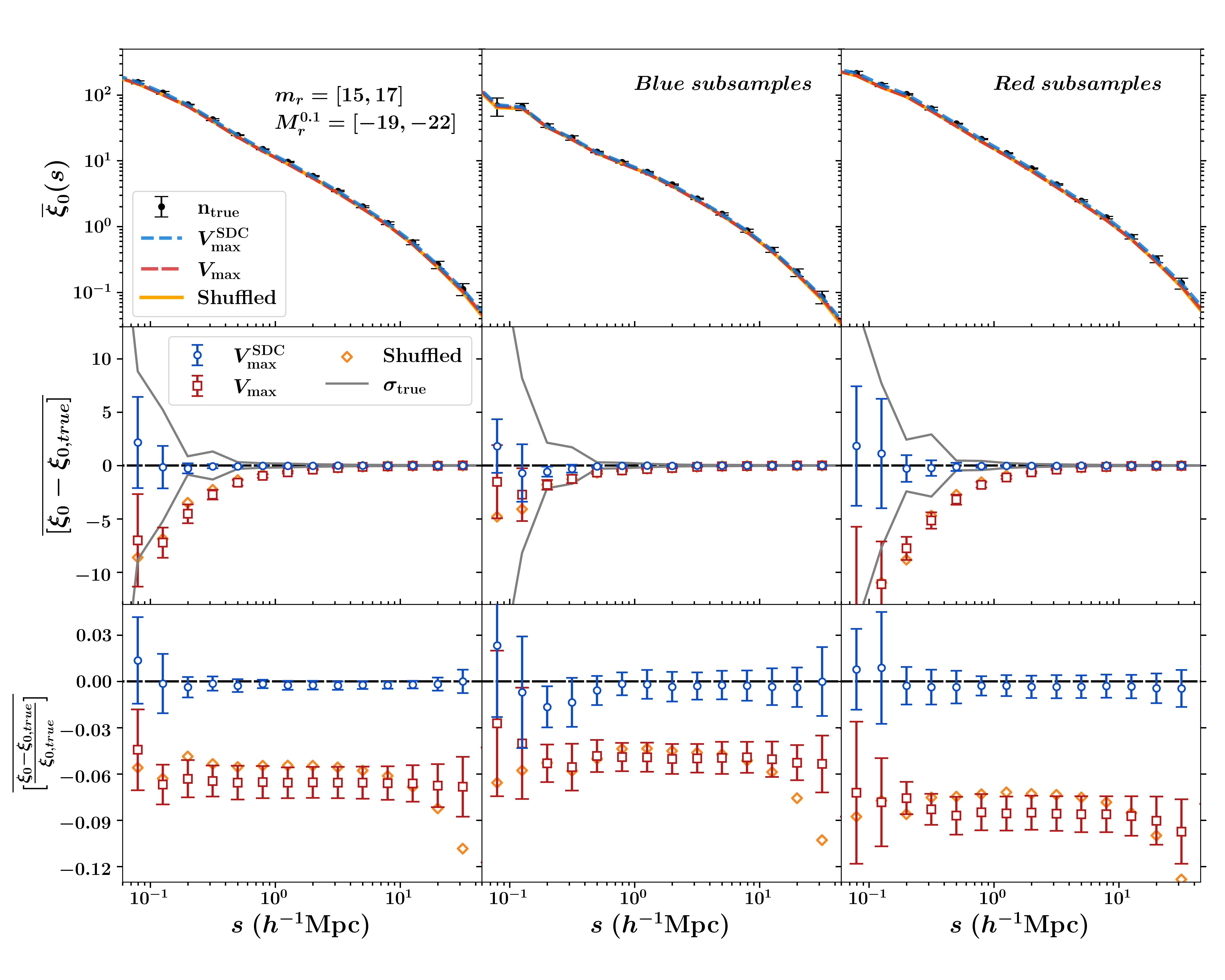}
  \caption{Similar to Figure~\ref{fig:wp_LS1}, with a comparison of the $\xi_0$ 
  for the redshift-space 2PCFs of the LS1 samples (left panels) and their blue 
  (middle panels) and red (right panels) subsamples.}
\label{fig:xi0_LS1}
\end{center}
\end{figure*}

\begin{figure*}
\begin{center}
\centering
  \epsscale{1.1}
  \plotone{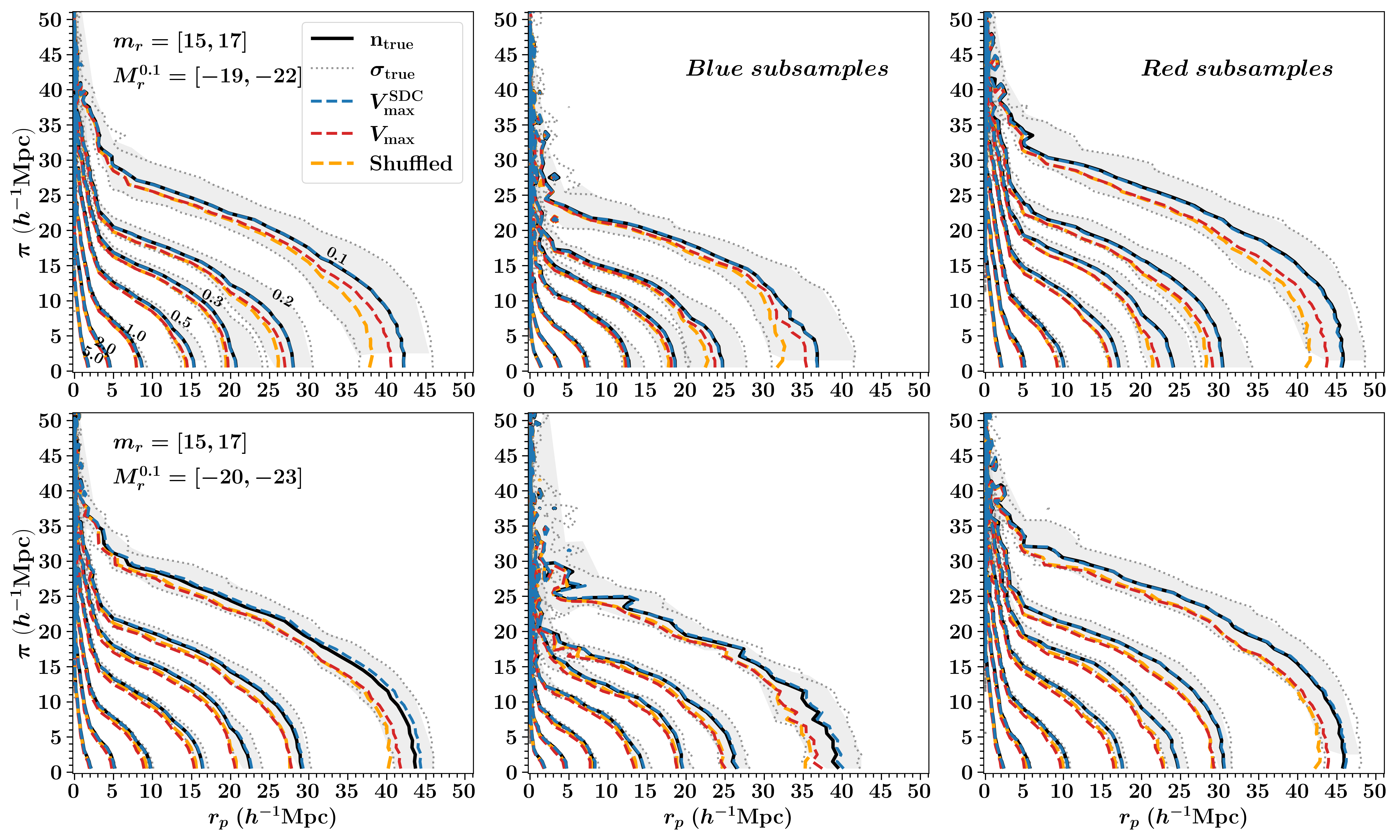}
  \caption{Comparison of the average 2D correlation function $\overline{\xi}(r_p,\pi)$ for 
  the luminosity-dependent flux-limited samples. The L1-C1, L2-C1, 
  and L3-C1 samples are shown from top to bottom accordingly, their blue/red 
  subsamples are shown in the middle and right panels in each row. Here, 
  $\overline{\xi}(r_p,\pi)$ is the averaged $\xi(r_p,\pi)$ among the 60 mock samples. 
  The true $\overline{\xi}(r_p,\pi)$ measured using the random catalog from 
  the $n(z)_{\rm true}$ method is shown as the black contour. The gray-shaded region with 
  dotted lines marks the $1\sigma$ scatter of the true $\overline{\xi}(r_p,\pi)$ 
  among the 60 mock samples. The yellow, red, and blue-dashed contours denote 
  the $\overline{\xi}(r_p,\pi)$ of the shuffled, \vmax, 
  and \sdcvmax\ methods, respectively. The contour levels from outside-in 
  correspond to $\overline{\xi}(r_p,\pi)=[0.1, 0.2, 0.3, 0.5, 1.0, 2.0, 5.0]$. 
  The middle and right column panels show s comparison of 
  the blue/red subsamples}
\label{fig:2dcf_LS}
\end{center}
\end{figure*}

The redshift-space correlation functions are compared in the same manner as 
$w_p$ for both the LC and LS samples, and the results for different radial selection 
models are generally consistent with the comparisons for $w_p$ in the previous section. 
The mean $\overline{\xi}_0$, $\overline{[\xi_0-\xi_{0,true}]}$, and 
$\overline{[(\xi_0-\xi_{0,true})/\xi_{0,true}]}$ for the LC samples 
with simple $k+e$ corrections are shown in Figure~\ref{fig:xi0_kpluse}, 
from top to bottom, respectively. Estimates of $\xi_0$ derived from the random catalogs 
created by the \sdcvmax\ approach display the smallest offsets 
and deviations from $\xi_{0,true}$ for both the LC1 (left panels) and LC2 (right panels) samples. 
For the \dcvmax\ technique, the $\xi_0$ at scale $r_p < 1 \mpch$ exhibit large offsets 
and deviations compared to the findings of $w_p$. For the \vmax\ method, 
$\xi_0$ deviations are marginally attenuated compared to the results of $w_p$, 
indicating that the impact of density fluctuations on clustering is less significant 
in redshift space. The $\xi_0$ for the shuffled approach exhibit the same 
offsets and deviations from $\xi_{0,true}$ as $w_p$. 
As the results of the LC samples without $k+e$ corrections are similar to those shown in 
Figure~\ref{fig:xi0_kpluse}, they are omitted here.

Figure~\ref{fig:xi0_LS1} illustrates the comparison of $\xi_0$ for the LS1 samples (left panels), 
and their blue (middle panels) and red (right panels) subsamples, respectively. 
Compared to the \vmax\ and shuffled methods, the \sdcvmax\ approach produces the least 
offsets and deviations from $\xi_{0,true}$ for the LS1 samples and red subsamples. 
For the blue subsamples, the \sdcvmax\ method's mean offset at $s \sim 0.07\mpch$ 
is slightly larger than the \vmax\ method's mean offset, and both approaches have comparable 
deviations at that scale. This is not a concern because the amount of uncertainty at this scale 
is also high due to the shot noise. In general, on $\xi_0$ measurements, the \sdcvmax\ technique 
continues to outperform the other two radial selection models. 
Since the findings of the LS2 samples are basically consistent with Figure~\ref{fig:xi0_LS1}, 
they are also excluded here.

In Figure~\ref{fig:2dcf_LS}, the average 2D correlation functions $\overline{\xi}(r_p,\pi)$ 
for the LS samples are presented. The $\overline{\xi}(r_p,\pi)$ for the LS1 samples (left panel), 
blue subsamples (middle panel), and red subsamples (right panel) are displayed in 
the upper panels. $\overline{\xi}(r_p,\pi)$s for the $n_{true}$, 
\sdcvmax, \vmax, and shuffled methods are 
represented by black solid lines, blue-dashed lines, red-dashed lines, 
and yellow-dashed lines, respectively. The $1\sigma_{true}$  dispersion of 
$\xi_{true}(r_p,\pi)$ among the 10 mock samples is denoted by gray-dotted lines 
in places with shading. The $\overline{\xi}(r_p,\pi)$ of the \sdcvmax\ model provide 
the best agreement with $\overline{\xi}_{true}(r_p,\pi)$ for the LS1 samples 
and color-dependent subsamples. For $\overline{\xi}(r_p,\pi)$ of the \vmax\ 
and shuffled methods, there are offsets of varying degrees; yet, the offsets stay 
within the $1\sigma_{true}$ error margins; however, the contour shapes are altered. 
In the lower panels displaying $\overline{\xi}(r_p,\pi)$s for the LS2 samples, 
the majority of contours for the \sdcvmax\ model are consistent with 
$\overline{\xi}_{true}(r_p,\pi)$. $1\%\sim 2\%$ deviations are seen in 
$\overline{w}_p$ (bottom left panel in Figure~\ref{fig:wp_LS2}) for both LS2 
samples and blue subsamples are also observed in contours at large scale. 
For the \vmax\ and shuffled methods, 
the offsets in the $\overline{\xi}(r_p,\pi)$ contours are close to the error 
margins of $1\sigma_{true}$; thus, the contour shapes are altered as well. 
Since the comparisons for the LC samples are substantially identical to those in 
Figure~\ref{fig:2dcf_LS}, they are excluded here.

\section{Discussion} \label{sec:disc}

Our tests demonstrate that for flux-limited sample with 
a redshift-dependent number density $n(z)$, utilizing the random catalog 
generated by the \sdcvmax\ technique to measure galaxy clustering produces 
the least deviation from the true clustering when compared to the other 
radial selection methods. Some aspects of the performance 
of the \sdcvmax\ technique remain to be clarified and discussed, as detailed below.


\subsection{Impact of Smoothness Parameters on Clustering Estimation} \label{sec:dis_smooth}

\begin{figure*}
\begin{center}
\centering
  \epsscale{1.}
  \plotone{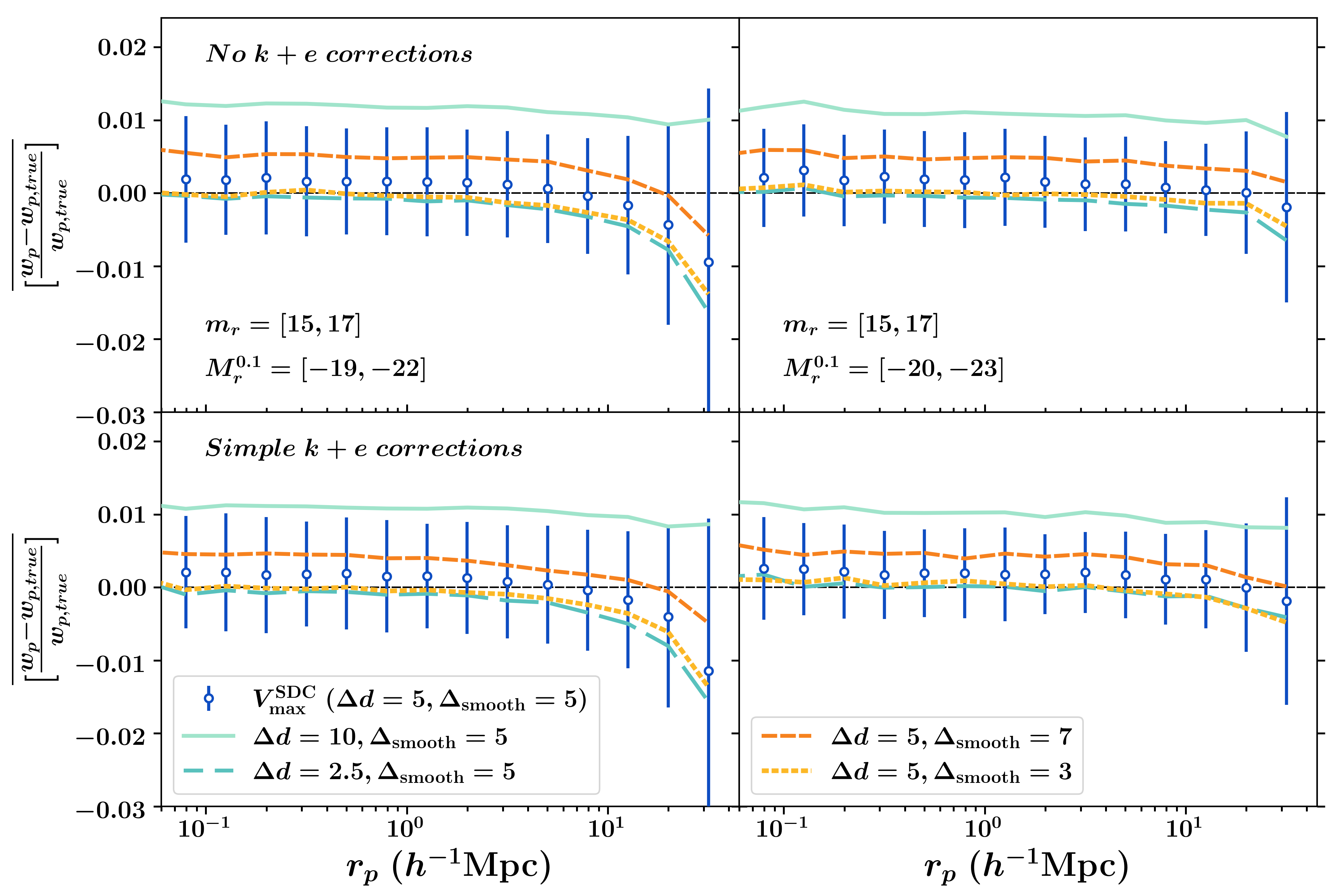}
  \caption{The average deviations of $w_p$ from $w_{p,true}$ 
  for the \sdcvmax\ method, in which alternative histogram bin sizes and 
  smooth box sizes are adopted in the smooth process in order to assess 
  the impact of multiple choices on clustering estimation. The fiducial bin size 
  and smooth box size used in Section~\ref{sec:comparison} are 
  $\Delta d=5 \mpch$ and $\Delta_{\rm smooth}=5$, respectively, 
  as indicated by the open blue circles with error bars.
  The alternate histogram bin sizes are $\Delta d=2.5 \mpch$ 
  and $\Delta d=10 \mpch$, with the same smooth box size 
  as the fiducial one, as indicated by the green-dashed lines and the light blue lines, 
  respectively. The alternate smooth box sizes are $\Delta_{\rm smooth}=$ 3 
  and 7, with the same fixed histogram bin size as 
  the fiducial one, as shown by the yellow short-dashed and 
  orange long-dashed lines, respectively. 
  The zero deviation is shown by the horizontal black-dashed lines.
  Upper panels: tests for the LC1 samples (left panel) and LC2 samples 
  (right panel) for the no $k+e$ correction case. 
  Lower panels: similar tests for LC1 and LC2 samples 
  to those in the upper panels, but for the simple $k+e$ correction case.}
\label{fig:comp_smooth}
\end{center}
\end{figure*}


For the \sdcvmax\ approach, we add a smoothing step to eliminate 
the unanticipated small fluctuations in the redshift distribution of 
the cloned random galaxies generated by the \dcvmax\ method. 
The previous comparison of 2PCFs for the \sdcvmax\ and 
\dcvmax\ methods demonstrate the necessity of a smoothing procedure 
for a random catalog in order to produce a nearly unbiased clustering 
measurement for the flux-limited sample.
Smoothing requires a selection of histogram bin size $\Delta d$ 
and smooth box size $\Delta_{\rm smooth}$. 
To determine the effect of varying $\Delta d$ and $\Delta_{\rm smooth}$ 
values on the final galaxy clustering determination, we vary these two 
smoothness parameters and regenerate random catalogs to perform the estimate. 
First, we set $\Delta d=5 \mpch$ and $\Delta_{\rm smooth}=5$ as the fiducial case, 
which we have used for the \sdcvmax\ technique in the previous tests 
in Section~\ref{sec:comparison}. Second, we chose $\Delta d=2.5 \mpch$ 
and $10 \mpch$ for the histogram bin size, with $\Delta_{\rm smooth}=5$ set to smooth. 
Thirdly, we select $\Delta_{\rm smooth}=3$ and $7$ for the smoothing with 
$\Delta d=5 \mpch$ set. Figure~\ref{fig:comp_smooth} displays 
the average deviations of $w_p$ from $w_{p,true}$ 
for the random catalogs created by the \sdcvmax\ technique 
with various $\Delta d$ and $\Delta_{\rm smooth}$ values. 
To simplify the assessment, we just test the projected 2PCFs 
of the LC samples here. In the absence of $k+e$ corrections, 
the upper panels of Figure~\ref{fig:comp_smooth} depict 
the mean deviations of $w_p$ for the LC1 (left panel) 
and LC2 (right panel) samples, respectively. 
We can see that a finer value of $\Delta d=2.5 \mpch$ (green-dashed lines) 
and $\Delta_{\rm smooth}=3$ (light blue lines)
lead to a constant drop in $\overline{[(w_p-w_{p,true})/w_{p,true}]}$ 
on all test scales, resulting in reduced deviations at $r_p \lesssim 2\mpch$ 
and an underestimate on a larger scale, especially for the LC1 samples.
In contrast, a coarser size of $\Delta_{\rm smooth}=7$ (orange long-dashed lines) 
results in an overall increase relative to the mean deviation in the fiducial case 
(open blue rolls with error bars), resulting in an overestimation at scale 
$r_p  \lesssim 20 \mpch$. A coarser size of $\Delta d = 10 \mpch$ 
(yellow short-dashed lines) leads to an $\sim 1\%$ increase 
in the mean deviation of $w_p$ relative to 
the deviation in the fiducial case; this is the only mean deviation that exceeds 
the $1\sigma$ errors but is still around $\sim 1\%$. 
In the lower panels, the test results for the LC samples with simple $k+e$ 
corrections are displayed, which are essentially identical to 
the findings in the above panels, suggesting that the smooth process 
is insensitive to galaxy samples when different $k+e$ corrections are applied.
 Our tests indicate that the variation in $\Delta d$ and $\Delta_{\rm smooth}$ 
 in the smooth process of the \sdcvmax\ technique affects the accuracy 
 of clustering measurement; however, the effect of deviations is much less than 1\%.
The advantage of the \sdcvmax\ technique over the other radial selection models still stands.

\subsection{Difference in Clustering Uncertainty} \label{sec:dis_samples}

\begin{figure}
\begin{center}
\centering
  \epsscale{1.2}
  \plotone{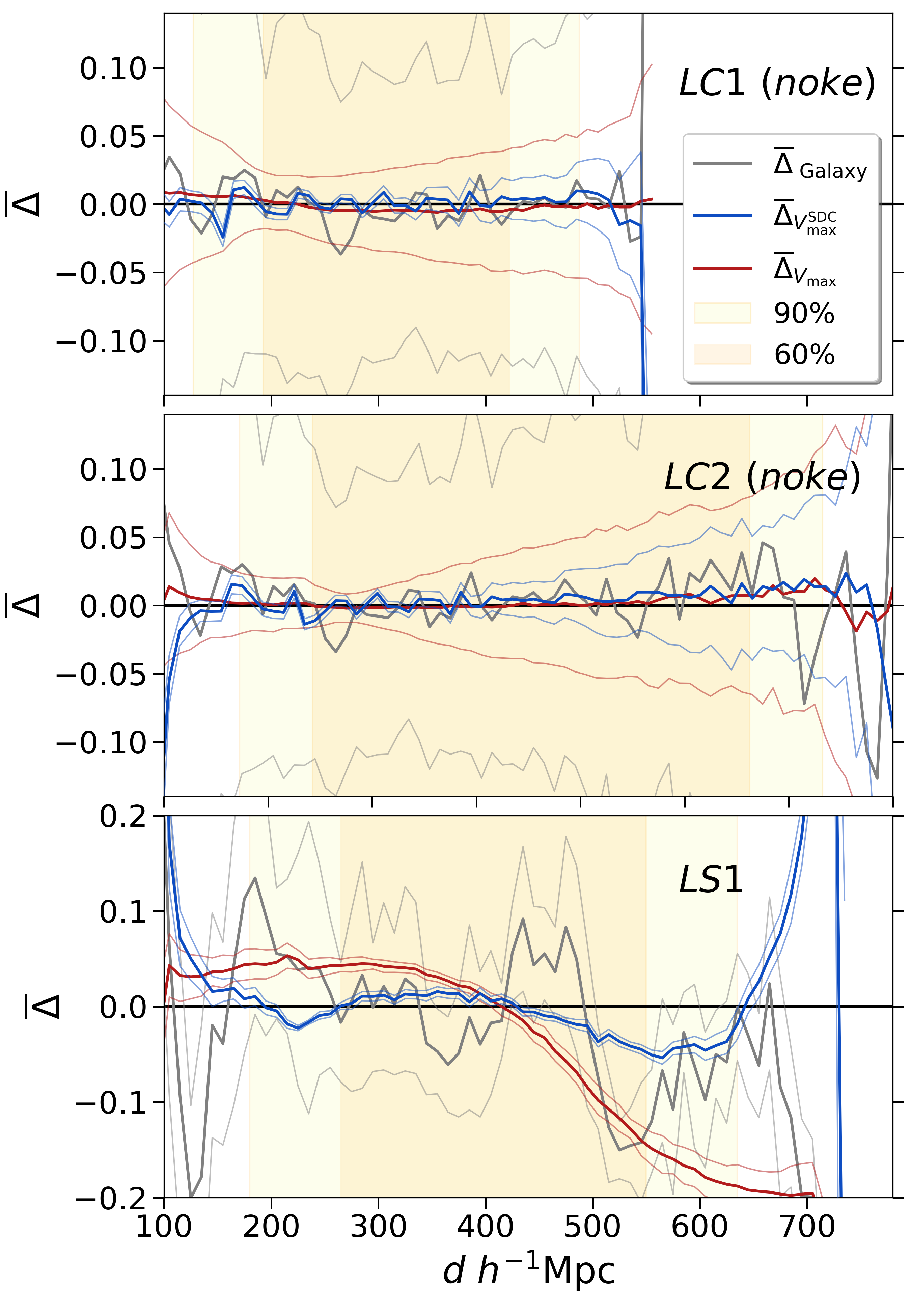}
  \caption{Top panel: the average deviations $\overline{\Delta}$ and $1\sigma$ errors 
  from the radial distribution of the random catalog obtained by the $n(z)_{\rm true}$ method. 
  The mean deviation is computed using the equation 
  $\overline{\Delta} = \overline{(n^i-n^i_{true})/n^i_{true}}$, where $n^i$ 
  is the normalized radial distribution of the $i$th LC1 sample and random catalog 
  produced using the \sdcvmax\ and \vmax\ methods. 
  The $\overline{\Delta}$ of the LC1 samples is shown by the thick gray lines, 
  while the $1\sigma$ errors over the 60 samples are represented by the thin gray lines. 
  The thick blue lines and thin light blue lines represent $\overline{\Delta}$ and errors 
  for the random catalogs generated by the \sdcvmax\ technique. 
  The thick red lines and thin light red lines represent the \vmax\ algorithm. 
  The light yellow and light orange regions indicate the locations of 90\% and 60\% 
  of galaxies, respectively. Middle panel: Similar to the top panel, it presents 
  the average deviations and errors for the LC2 samples and their corresponding 
  random catalogs. Bottom panel: Similar to top panel, it displays 
  the average deviations and errors for the LS1 samples and their corresponding random 
  catalogs.}
\label{fig:hist_bias}
\end{center}
\end{figure}

In prior tests, the uncertainties in clustering deviations among the 60 LC 
samples are significantly larger than the uncertainties in the 10 LS samples, 
which is not expected intuitively. In addition, the deviation uncertainties for 
the \sdcvmax\ approach are approximately a fourth of those for the \vmax\ method 
in the LC samples. As can be seen in Figure~\ref{fig:hist_bias}, we further investigate 
the radial distribution of the LC and LS samples in order to determine the probable 
distinct drivers of these discrepancies. Here, we take into account the LC samples 
without $k+e$ corrections and the LS1 samples, which are sufficient to explain 
the difference in uncertainty. First, we compute the normalized radial distribution 
for the galaxy samples and random catalogs created using the $n(z)_{\rm true}$, 
\sdcvmax, and \vmax\ methods, respectively. 
To quantify the density fluctuations relative to the true smooth distribution created 
by the $n(z)_{\rm true}$ method, we estimate the average deviations $\overline{\Delta}$ 
and $1\sigma$ variances of these distributions from the genuine normalized distribution 
for the 60 LC samples and 10 LS1 samples separately, as shown in Figure~\ref{fig:hist_bias} 
from top to bottom.

The $\overline{\Delta}$ and $1\sigma$ variance for the galaxy samples are shown 
by the thick gray and thin light gray lines. For both the LC1 (upper panel) and 
LC2 (middle panel) samples, the variations across the 60 individual samples vary greatly, 
as indicated by $1\sigma$ variance, whereas $\overline{\Delta}$ exhibits a relatively small 
deviation from the true normalized distribution. The light yellow and light orange regions denote the locations 
in which 90\% and 60\% of the expected random galaxies are likely to be distributed, 
and we anticipate that the bulk of pairs used to estimate clustering are from 90\% region. 
$\overline{\Delta}$ (red thick lines) and $\sigma$ (light red thin lines) of the \vmax\ technique 
reveal that this approach corrects the fluctuations in the galaxy samples; nonetheless, 
the imprints of large-scale structures are still discernible. 
For instance, $\overline{\Delta}$ for the LC1 samples shows a small but observable deviation 
at $100\sim 450 \mpch$ where 90\% of galaxies are located. This explains the consistent bias 
noticed in $w_p$ and $\xi$ in previous testing. For the LC2 samples, the systematic bias is almost imperceptible, 
with just a tiny overestimation at $d\gtrsim 500\mpch$, indicating a clustering bias that has been 
detected in prior tests. For the \sdcvmax\ approach, there are noisy fluctuations in 
$\overline{\Delta}$ (blue thick lines) for both the LC1 and LC2 samples, indicating that the smoothing 
does not eliminate all noisy fluctuations in the radial distribution and there is still room to improve the 
smoothing. Fortunately, these fluctuations are complimentary to a certain degree, 
yielding a substantially unbiased measurement for galaxy clustering. 
We observe that the $1\sigma$ errors (light blue thin lines) for the \sdcvmax\ approach 
are less than those for the \vmax\ method, especially for the LC1 samples at the 60\% region. 
This is essentially the reason for the substantial difference in uncertainty found between 
the two techniques for $w_p$ and $\xi$, demonstrating once again that the \sdcvmax\ method 
can more successfully rectify the effect of density fluctuations on individual samples, 
and thus the clustering estimations converge to the genuine galaxy clustering.

As demonstrated in the bottom panel, $\overline{\Delta}$ for the LS1 samples deviates 
significantly from the genuine distribution when compared to the LC samples. 
By rotating the sky, just 10 LS1 samples are created from a single light cone catalog. 
These samples have a significantly reduced $1\sigma$ variance than LC samples, particularly 
at the $60\%$ region. In the LS1 samples, the advantage of density correction in the \sdcvmax\ approach 
is exhibited more clearly compared to the \vmax\ method. Both approaches have equal errors, 
but the $\overline{\Delta}$ of the \sdcvmax\ method deviates less from the true distribution, 
resulting in a more accurate clustering measurement. In contrast, the \vmax\ technique 
predicts too many random galaxies at $d \lesssim 400$ and fewer galaxies at high $d$ 
due to strong fluctuations in galaxy samples, hence exhibiting a greater deviation 
in $\overline{\Delta}$ in comparison to $\overline{\Delta}$ of the LC samples. 
This also explains the extremely systematic bias in $w_p$ observed for 
the \vmax\ approach on all testing scales in earlier tests.

Last but not least, the LC samples and LS samples are derived from distinct 
parent mock catalogs utilizing two simulations with different resolutions and 
galaxy-halo connection models. Both the LC and LS samples are complete 
at $M^{0.1}_{\rm r} \leq -18$; however, the simulation of \citep{2019SCPMA..6219511J} 
used to generate the LC samples has a mass resolution that 
is an order of magnitude higher than that of the MXXL simulation \citep{2012MNRAS.426.2046A}, 
implying that more halo and galaxy structures are resolved in the LC samples. 
Moreover, despite the fact that the LC samples are constructed using a simple 
galaxy-halo connection model with simple $k+e$ corrections, 
the benefit is that all model parameters are clear and straightforward; 
hence, the potential deviation and error sources are comprehendible. 
For the LS samples, with a more sophisticated galaxy evolution 
and $k$ correction, the light cone catalog of \cite{2017MNRAS.470.4646S} 
is theoretically closer to actual observation data; the main drawback is a restricted 
number of samples. 
The test results of these two sample groups demonstrate that either 
the $k+e$ corrections are based on simple or more complex and realistic mock catalogs, 
the \vmax\ technique may produce an inaccurate measurement of galaxy clustering, 
whereas the \sdcvmax\ method can always produce an accurate and precise estimate of clustering.

\section{Conclusions} \label{sec:concls}

In this paper, we provide a radial selection model, the \sdcvmax\ approach, 
for generating the redshifts of random catalogs in galaxy two-point statistics 
that allows for a high level of accuracy and precision in the estimation.
This method is an improvement 
on the density-corrected \vmax\ method proposed by \cite{2011MNRAS.416..739C}, 
and it consists mostly of three modifications: (1) adding an estimate of $z_{\rm min}$ 
and expanding the code's application to a general flux-limited sample; 
(2) support for a redshift and color-dependent $k-$correction model applicable to individual galaxies; 
(3) adding a smooth step to the output cloned radial distribution of random galaxies.
These modifications are crucial for obtaining a smooth radial distribution for a random 
catalog that is unaffected by galaxy density fluctuations, which is the key to 
a clustering measure with high precision and accuracy.

We measure 2PCFs using two groups of flux-limited samples, designated LC and LS, 
to validate the \sdcvmax\ approach. The flux-limited LC samples are constructed 
from 10 mock catalogs with two luminosity cuts and two simple $k+e$ correction cases. 
Using the same sample selection criteria and luminosity thresholds as for the LC samples, 
t10 LS samples are generated using the light cone catalog of \cite{2017MNRAS.470.4646S}. 
To test property-dependent clustering, the LS samples are subdivided into blue and red subsamples. 
We compare the projected and redshift-space 2PCFs using random catalogs created from 
the $n_{true}$, \sdcvmax, \dcvmax, \vmax, 
and redshift shuffled methods. Our test results demonstrate that the \sdcvmax\ approach 
is the only reliable radial selection model capable of achieving sub-percent accuracy for 
$w_p$ measurement on scales ranging from $0.07\mpch$ to $\sim 40\mpch$. 
A $2\%$ deviation arises on a large scale for the LS2 sample; 
however, it is still less than the deviations of other radial selection models.
In general, the \sdcvmax\ technique can constrain the measurement accuracy of $w_p$ to 
within $1\%$ for color-dependent galaxy clustering, validating its superiority over 
the \vmax\ and redshift shuffled methods.

The next generation of spectroscopic surveys, specifically the DESI experiment, 
will obtain the spectra of around 40 million galaxies and quasars over 14,000 $deg^2$, 
which is almost an order of magnitude more than the previously observed galaxies 
\citep{2022arXiv220808518M}. These extragalactic objects include 13 million 
bright galaxy sample (a magnitude of 2 deeper than the SDSS main sample) 
\citep{2023ApJ...943...68L}, 8 million luminous red galaxies, 
16 million emission line galaxies, and 
3 million quasars \citep{2013arXiv1308.0847L,2016arXiv161100036D,
2016arXiv161100037D,2022arXiv220808513R}. 
On the one hand, the two-point statistics of these up-coming galaxies will surely 
afford us an unprecedented opportunity to comprehend the physics of galaxy 
formation and evolution, improve the galaxy-halo connection, and shed light on the role of 
the halo environment in determining the galaxy's physical properties \citep{2022ApJ...938L...2F}. 
On the other hand, how to fully exploit these galaxies, particularly with 
the assistance of the galaxy 2PCFs, remains a challenge. 
Using volume-limited catalogs to conduct the 2PCF analysis will not only 
result in the rejection of a considerable number of galaxies, 
but it may also lead to the loss of crucial information imprinted in clustering. 
The density-corrected \vmax\ approach proposed by \citep{2011MNRAS.416..739C} 
solves this problem, and our improvements and tests confirm that the \sdcvmax\ method 
is a viable technique for accurately measuring clustering for flux-limited and color-dependent samples, 
hence maximizing the use of galaxies.
Our present tests are preliminary, concentrating mostly on low-redshift galaxies. 
In the future, we will continue to improve this approach and conduct more tests on various 
properties of galaxies (e.g., stellar mass, star-formation rate, and so forth) 
as well as tests employing relative high-redshift galaxies (e.g., CMASS, BOSS and eBOSS) and mocks.

\begin{acknowledgments}

We appreciate the referee's insightful comments and suggestions, which substantially improve this article. 
We would like to thank Yipeng Jing for carefully reading the manuscript and providing valuable comments.
We are also grateful to Yipeng Jing for generously providing the simulation data.
Lei Yang expresses gratitude to Chun Xia for assisting with the use of 
the Yunnan University Astronomy Supercomputer.
This work is sponsored by grants from Yunnan University's Launching Research Fund 
for Postdoctoral Fellow (C176220200) and the China Postdoctoral Science Foundation (2020M683387).
The majority of calculations were performed on the Yunnan University Astronomy Supercomputer.

\end{acknowledgments}

\newpage

\appendix

The appendix contains supplementary information on the mock samples. 
Figure~\ref{fig:nd_mocks} displays the number density of galaxy samples as a function of redshift. 
Figure~\ref{fig:gsample} shows the galaxy distribution on a color-magnitude diagram for the LS samples.

\section{Mock Samples} \label{sec:append_nd}

\begin{figure*}
\begin{center}
\centering
  \epsscale{.6}
  \plotone{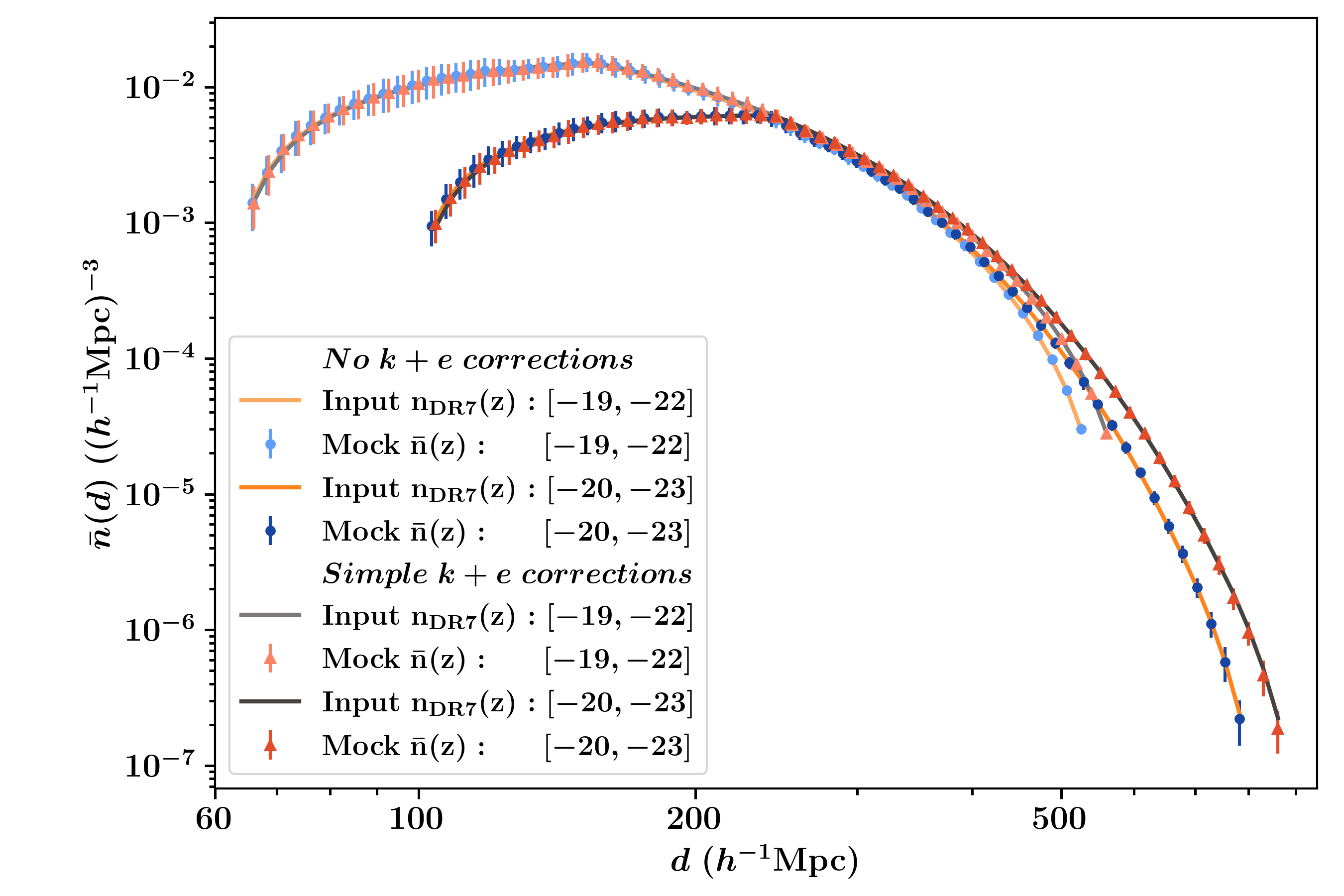}
  \caption{The mean number density $\overline{n}(z)$ among the 60 LC mock samples. 
  These samples have a flux cut at $m_r=[15, 17]$ and two luminosity cuts at 
  $M^{0.1}_r=[-19,-22]$ for the LC1 samples and $M^{0.1}_r=[-20,-23]$ for the LC2 samples. 
  In the case of no $k+e$ corrections, the light blue and dark blue points with error bars 
  represent the $\overline{n}(z)$ and $1\sigma$ variance for the LC1 and LC2 samples, 
  respectively. The orange and light orange lines show the input $n_{\rm DR7}$ derived by 
  the input LF and sample selection criteria. In the case of simple $k+e$ corrections, 
  the orange and red triangles with errors indicate $\overline{n}(z)$ and $1\sigma$ 
  for the LC1 and LC2 samples, respectively. The inputs $n_{\rm DR7}$ are shown in gray 
  and light gray lines.}
\label{fig:nd_mocks}
\end{center}
\end{figure*}

As an example, Figure~\ref{fig:nd_mocks} displays the estimated average 
galaxy number density $\overline{n}(z)$ for the 60 LC samples. 
The $\overline{n}(z)$ of these flux-limited samples changes as a function 
of comoving distance. The $\overline{n}(z)$s of the LC samples 
are in excellent agreement with the predicted input $n_{\rm DR7}$ derived from 
the input LF and the corresponding sample selection criteria. 
As predicted, $\overline{n}(z)$ for the LS2 samples contains more brighter and 
high-redshift galaxies than $\overline{n}(z)$ for the LS1 samples. 
In addition, the $\overline{n}(z)$ for the samples with simple $k+e$ corrections 
exhibits a slight evolution toward higher redshift when compared to samples 
without $k+e$ corrections.

\begin{figure*} 
   \centering
   \includegraphics[width=15.0cm, angle=0,scale=.8]{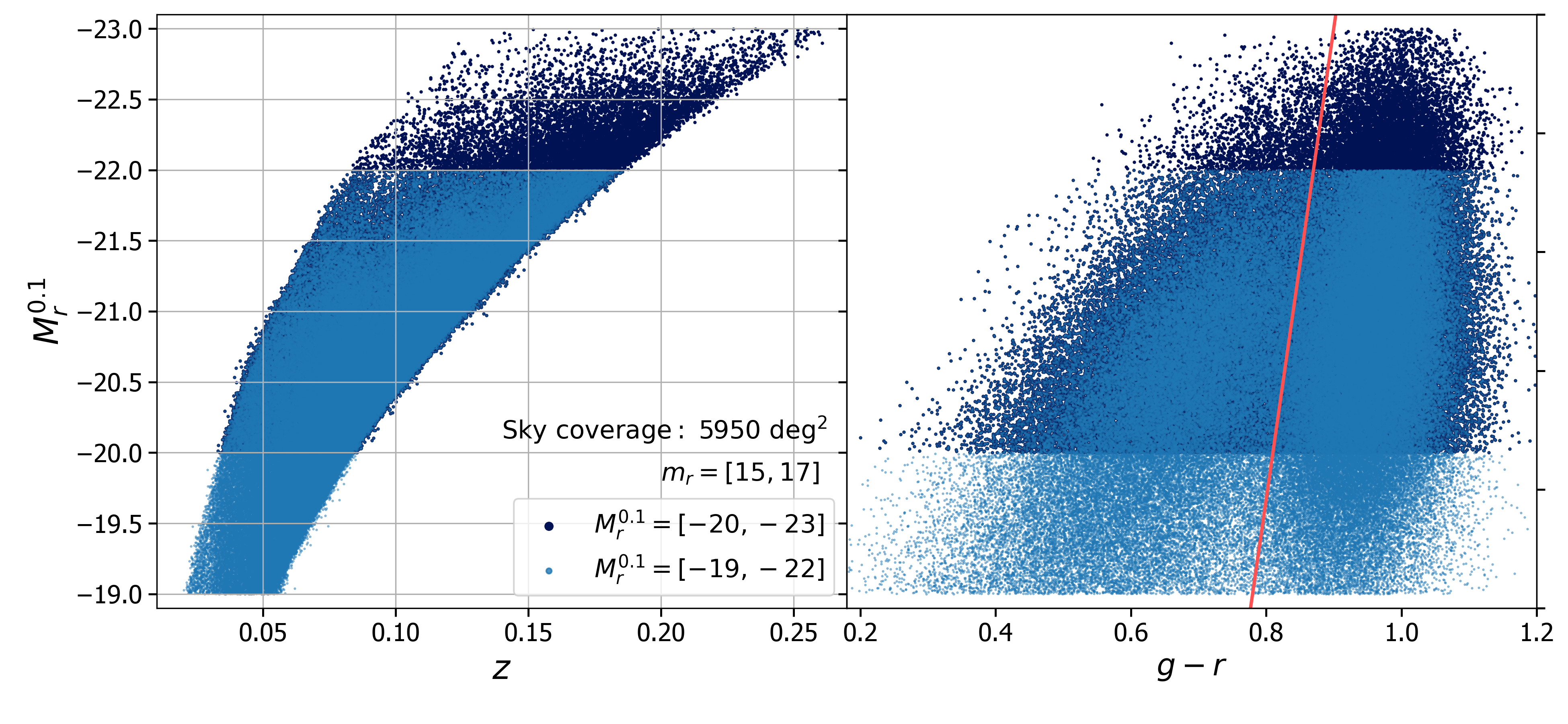}
   \caption{Left panel: the LS samples on the magnitude-redshift diagram. 
   The light blue points denote galaxies in one of the LS1 samples with a flux 
   cut at $m_r=[15,17]$ and luminosity cut at $M^{0.1}_{\rm r}=[-19,-22]$. 
   The dark blue points stand for one of the LS2 samples with cuts of 
   $m_r=[15,17]$ and $M^{0.1}_{\rm r}=[-20,-23]$. The samples have a 
   sky coverage of $5950\deg^2$. Right panel: the LS samples on the 
   color-magnitude diagram. The LS1 and LS2 samples are color coded similarly 
   to those in the panel on the left. The red line as referred to by \cite{2011ApJ...736...59Z} 
   splits galaxies into blue and red subsamples.}
   \label{fig:gsample}
\end{figure*}

Figure~\ref{fig:gsample} displays the LS samples on the redshift-magnitude diagram 
(left panel) and color-magnitude diagram (right panel), respectively. The flux-limited 
LS samples are constructed from a light cone catalog with two luminosity cuts. 
At the low-redshift regions, the light cone catalog  mimics the SDSS DR7 data and hence, 
has an LF of \cite{2003ApJ...592..819B}. We use the method described in 
\cite{2011ApJ...736...59Z} to divide the galaxies into blue and red galaxies, 
as indicated by the red line in the right panel. Additionally, the LS samples 
have a redshift-dependent number density identical to that observed in 
Figure~\ref{fig:nd_mocks} and spanning a broader redshift range.

\bibliography{smoothvmax}{}
\bibliographystyle{aasjournal}


\end{CJK}

\end{document}